

\def\Reals{{\bf R}}
\def\Ints{{\bf Z}}
\def\Complexs{{\bf C}}
\def\Naturals{{\bf N}}
\def\oj{{\bf g}}

\def\one{{\bf 1}}

\def\LL{{\cal L}}
\def\TT{{\bf T}}
\def\CC{{\bf C}}

\def\dn{{\tilde d}}

\def\gamman{{\tilde \gamma}}

\def\pr{^\prime}

\def\lc{{\lambda \over c}}

\def\forall{\hbox{for all }}

\def\tr{\hbox{\rm tr }}

\def\IX{\hbox{\raise.23ex\hbox{{$\scriptscriptstyle\mid$}}
\kern-0.62em\hbox{{$\times$}}}}

\def\d{\partial}
\def\ddot{{\kern-0.07em\cdot\kern-0.07em}}

\def\e{\hbox{e}}
\def\emx{\e^{m \ddot x}}
\def\enx{\e^{n \ddot x}}
\def\emnx{\e^{(m+n) \ddot x}}

\def\args#1#2{{{#1_1} \ldots {#1_{#2}}}}
\def\argsx#1#2#3{{{#1_1} \ldots #3 \ldots {#1_{#2}}}}

\def\state#1{{\kern 0.1em | \kern-0.05em #1 \kern-0.05em \rangle}}
\def\vac{\state 0}
\def\hvac{\state h}

\def\qvac{\state q}

\def\bP{{\bf P}}

\def\nup{\args \nu p}

\def\tauq{\args \tau q}

\def\xxx{\leftrightarrow}
\def\mxn{m \xxx n}

\def\psic{\psi^\dagger}
\def\Gammac{\Gamma^\dagger}

\def\point{{$\bullet\;$}}

\def\half{{1 \over 2}}

\def\mapright#1{\hbox{$\smash{\mathop{\longrightarrow}\limits^{#1}}$}}

\def\refto#1{{${}^{#1}$}}
\def\header#1{ {\bf #1} \nobreak}

\def\endchapter{{\vskip 15 mm} \goodbreak}
\def\newpage{{\vfill \eject}}
\baselineskip = 12pt

\overfullrule = 0pt
\magnification = 1200

\centerline{\bf Vect(N) Invariants and Quantum Gravity }

\vglue 30pt

T. A. Larsson\footnote*{Supported by the Swedish Natural
Science Research Council (NFR)}

Department of Theoretical Physics,

Royal Institute of Technology

100 44 Stockholm, Sweden

email: tl@theophys.kth.se

\vglue 30pt
(April 1992)
\vglue 60pt

\header{Abstract}

The Feigin-Fuks construction of irreducible lowest-weight Virasoro
representations is reviewed using physics terminology. The procedure
consists of two steps: constructing invariants and applying them to
the Fock vacuum. We attempt to generalize this construction to the
diffeomorphism algebra in higher dimensions. The first step is
straightforward, but the second is difficult, due to the appearence of
infinite Schwinger terms.  This might be avoided by imposing
constraints on the fields, which should be of the recently discovered
conformal type. The resulting representations are reminiscent of
quantum gravity.

\vglue 2cm

\vglue 1cm
hep-th/9209092

\newpage

\header{1. Introduction}

Conformal field theory applied to two-dimensional phase transitions is
one of the most powerful theories in physics.\refto{1,2} A striking
success is the explanation of universality, i.e. the phenomenon that
similar but different models have exactly the same critical exponents.
Namely, because the simplest consistent continuum theories fall into a
discrete series, a small change in the model must either result in a
large change in the universal behavior, or in no change at all.

Universality is seen experimentally also in three dimensions, and it
is built into the renormalization group approach to critical
phenomena.\refto{3} It would therefore be desirable to have a theory
with the same predictive power as conformal field theory also in
higher dimensions. At first sight, this appears impossible, because
the conformal group in two dimensions is intimately tied to the
existence of complex numbers, which are inherently two-dimensional.
However, if the relevent group is enlarged already in two dimensions,
this larger group may have a generalization to higher dimensions.
There is only one viable candidate, namely the diffeomorphism group in
$N$ dimensions.

It can also be argued on purely mathematical grounds that any sensible
object transforms as a representation of the diffeomorphism group,
i.e. that it is generally covariant. From a passive point of view, a
diffeomorphism is simply a coordinate transformation, so
representations of the diffeomorphism group have a meaning
irrespective of the choice of coordinate system. In the language of
differential geometry, they are intrinsic objects. In fact,
differential geometry\refto{4,5} deals with the properties of such
objects, and consequently that entire subject can be described as
diffeomorphism group representation theory.

The diffeomorphism group is also essential to quantum gravity, in
approximately the same way as $SU(2)$ is important to the quantum
theory of spin.  Its representation theory thus seems to be an
important subject to study, and it has recently attracted some
interest by physicists.\refto{6-12} For simplicity, we limit ourselves
to the infinitesimal form of the diffeomorphism group, i.e. the Lie
algebra of vector fields in $N$ dimensions, $Vect(N)$.  This notation
makes sense because we are only interested in local aspects; the only
important parameter is the dimension of space.

The basic question underlying this work is the following. How do
quantum fields transform under arbitrary coordinate transformations?
In one complex dimension, the answer is well known: quantum fields
transform as unitary, irreducible, lowest-weight representations of
the Virasoro algebra, and it therefore seems to be a reasonable
assuption that quantum fields in higher dimensions likewise transform
as unitary, irreducible, lowest-weight representations of $Vect(N)$,
extended by some Schwinger terms (anomalies).  In principle, this
statement would amount to a classification of inequivalent quantum
fields. Since critical exponents arise as dilatation eigenvalues, and
a dilatation is certainly a diffeomorphism, we would also obtain a
classification of higher-dimensional critical exponents, which would
be very important in statistical physics.

Unfortunately, there is a slight complication. The classification of
the above-mentioned class of representations has not been achieved
except in one dimension, and we are in fact not aware of any
non-trivial higher-dimensional example.  In one real dimension, the
algebra has no non-trivial unitary representations, but if the
dimension is complex the situation is more interesting. We can use the
monomial basis
$$ L_m = z^{m+1} {d \over {dz}},
	\eqno(1.1) $$
which leads to the Witt algebra
$$ [L_m, L_n] = (n - m) L_{m+n}.
	\eqno(1.2) $$
However, we must recall that (1.2) arose as the infinitesimal form of
the diffeomorphism group close to the origin. If the generators should
act as infinitesimal transformations, they must in particular not be
infinite in some neighboorhood of the origin, including the origin
itself.  Hence any generator with a pole at the origin should be
discarded. From (1.1) it is then clear that relevant algebra must be
restricted to the following ``amputated'' Witt algebra (algebra of
{\it holomorphic} vector fields),
$$ [L_m, L_n] = (n - m) L_{m+n},
\qquad m, n \ge -1.
	\eqno(1.3) $$
The amputated algebra has many interesting representations, because it is
a subalgebra of the Virasoro algebra for any value of $c$.
In the Virasoro algebra,
$$ [L_m, L_n] = (n - m) L_{m+n} - {c \over {12}} (m^3 - m) \delta_{m+n},
	\eqno(1.4) $$
the central extension only appears in brackets where either $m$ or $n$
is less than $-2$, and hence any Virasoro representation yields a
representation of (1.3) by restriction.

An important technical point is that the Verma module must be
inverted, compared to the usual convention in the physics literature.
Let $\hvac$ be the state characterized by
$$ L_0 \hvac = h \hvac, \qquad L_{-1} \hvac = 0.
	\eqno(1.5) $$
The Verma module then has the basis
$$ \ldots {L_3}^{p_3} {L_2}^{p_2} {L_1}^{p_1} \hvac,
	\eqno(1.6) $$
with only finitely many $p_k$ non-zero.
This module can be extended to a Virasoro representation by
$$ L_{-m} \hvac = 0, \qquad \forall -m < 0.
	\eqno(1.7) $$
Although mathematically isomorphic, the inversion of the Verma module
has important physical consequences. Because $L_{-1}$ is the
translation operator, the ground state $\hvac$ is translationally
invariant. This is natural if we want to identify it with a state of
{\it uniform} incipient magnetization. On the other hand, $\hvac$ is
not conformally invariant, but it is conformally covariant by (1.6).
This should be contrasted to the opposite convention, which is
prevailing in physics, with a ground state with is not translationally
invariant. It is very difficult for us to picture such a ground state.

So where did the central extension go? The answer is that it
resurfaces when we consider the singular transformations $L_{m}$,
$m \le -2$. Classically, (1.3) can obviously extended to negative $m$
just be lifting the restriction, but in the quantum case a surprise
turns up in the form of a non-zero central extension.  Moreover, we
are really interested in unitary representations, which need an
involution to be defined. The natural definition $L^\dagger_m =
L_{-m}$ is consistent with the grading, but contrary to the situation
for finite-dimensional Lie algebras, there are many inequivalent ways
to define the bracket between the original and involuted generators.
These inequivalent involutions are parametrized by the value of $c$.
Similarly, we expect that Schwinger terms appear when we consider
singular transformations in higher dimensions.

The present work makes some progress towards irreducible
representations of $Vect(N)$, at least by generalizing half of
Feigin's and Fuks' construction of irreducible Virasoro
representations.\refto{13,2} As explained in Sec. 2, their work
proceeds in two steps: constructing invariants in an exterior algebra,
and relating them to lowest-weight modules in a Fock space. The first
step can readily be generalized to higher dimensions, which is done in
Sec. 4. The second step is more difficult, because the anomalies tend
to diverge, but we think that this problem can be resolved in a
multi-field Fock space by imposing constraints on the fields. Indeed,
all possible covariant equations have to be satisfied, or else the
representation is decomposable.  Some possible constraints on tensor
fields are listed in Sec. 5. In Sec. 6 a similar list is given for the
recently discovered {\it conformal fields},\refto{11} which in a sense
are more natural than tensor fields, and an algebraic structure not
completely dissimilar to canonical quantum gravity arises. Because
this structure appears in a mathematically very natural context,
namely pure representation theory, we suggest that we might have found
a viable approach to consistent quantum gravity. At least, we do know
that $Vect(N)$ plays an important role in quantum gravity, namely as
the symmetry algebra.

\endchapter

\header{2. Discrete Virasoro series }

The Witt algebra $Vect(1)$,
$$ [L_m, L_n] = (n - m) L_{m+n},
	\eqno(2.1) $$
can be realized as vector fields on the circle or on the line,
$$ L_m = -i \e^{imx} {d \over {d x}} = -i \e^{imx} \d,
	\eqno(2.2) $$
where $m \in \Ints$ on the circle and $m \in \Reals$ on the line. The
Witt algebra also has the realization (1.1), with $m \in \Ints$. Since we
are working on the level of linear representations in this paper, the
imaginary unit can be absorbed into the definition of $m$,
$$ L_m = \e^{mx} \d,
	\eqno(2.3) $$
which saves many explicit references to $i$.
All cases are summarized by $m \in \Lambda$, where $\Lambda$ is an abelian
group, possibly continuous or imaginary.

An important $Vect(1)$ module is the primary field $\TT(\lambda, w)$.
The action is given by
$$ [L_m, \psi(x)] = -\e^{mx}(\d + \lambda m + w) \psi(x)
	\eqno(2.4) $$
In the Fourier transformed basis, defined by
$$ \psi(x) = \sum_{n \in \Lambda} \psi_n \e^{-nx},
	\eqno(2.5) $$
(2.4) takes the form
$$ [L_m, \psi_n] = -(\d + \lambda m + w) \psi_{m+n} \equiv
(n + (1-\lambda) m - w) \psi_{m+n}
	\eqno(2.6) $$
where the derivative in momentum space simply is multiplication by a
constant: $\d \psi_n = -n \psi_n$, $\d \psi_{m+n} = -(m+n) \psi_{m+n}$.
The adjoint representation is clearly $\TT(2, 0)$.  The
parameter $w$ can be shifted to $w + s$ by relabelling the Fourier
components as $\psi_n \to \psi_{n+s}$.  It is thus only defined modulo
$\Lambda$, and particularly for the plane-wave basis on the line it
can be eliminated alltogether. However, on the circle and for the
basis (1.1), it is an important parameter.

Another natural module is the connection (one-dimensional
Christoffel symbol),
$$ [L_m, \Gamma(x)] = -\e^{mx} \big((\d + m)\Gamma(x) - m^2),
	\eqno(2.7) $$
transforming as the primary field $\TT(1, 0)$ apart from an
inhomogeneous term.
The covariant derivative is a map
$\TT(\lambda, w) \longrightarrow \TT(\lambda+1, w)$, which depends on
the connection. It is explicitly given by
$$ \nabla \psi = (\d + \lambda \Gamma + w) \psi
	\eqno(2.8) $$

It is not difficult to prove by direct computation that (2.7) is
consistent and that (2.8) defines a module map.

The above modules extend by means of Leibniz' rule to tensor products of
$\psi$ and $\Gamma$.
The field $\psi(x)$ can be either bosonic or fermionic;
$$ [\psi(x), \psi(y)] = 0,
\qquad \hbox{or} \qquad
\{\psi(x), \psi(y)\} = 0,
	\eqno(2.9) $$
respectively. The connection $\Gamma(x)$ must be bosonic, because the
inhomogeneous term in its transformation law is so.
A basis for these modules is given by all states of the form
$$ \psi(x_1) \ldots \psi(x_p) \Gamma(x_{p+1}) \ldots \Gamma(x_{p+q}),
	\eqno(2.10) $$
modulo the relations above.
The number operators $N_\psi$ and $N_\Gamma$ satisfy
$$ [N_\psi, \psi(x)] = \psi(x), \qquad
[N_\Gamma, \Gamma(x)] = \Gamma(x).
	\eqno(2.11) $$
$N_\psi$ commutes with every $L_m$ and thus the module (2.10)
decomposes into sectors with a fixed number of $\psi$'s. However,
$N_\Gamma$ does not commute with $Vect(1)$ due to the inhomogeneous
term in (2.7).

We will use a terminology inspired by physics and refer to (2.4) as
the transformation law of a {\it particle} and (2.10) as a state with
$p$ {\it fundamental particles} and $q$ {\it gauge bosons}.  It is
natural to try to construct a {\it molecule}, i.e. a composite
particle, in this multiparticle state, and to ask what the
corresponding values of $\lambda$ and $w$ are.

Because of locality, a molecule must be constructed out of $\psi(x)$,
$\Gamma(x)$, and finitely many of their derivatives at the same point
$x$. Using the fact that the covariant derivative maps primary fields
to primary fields, the most general local expression reads
$$ \Phi^\bP(x) \equiv \Phi^{(p_0, p_1, p_2, \ldots)}(x)
= \psi(x)^{p_0} (\nabla \psi(x))^{p_1} (\nabla^2 \psi(x))^{p_2} \ldots
	\eqno(2.12) $$
where only finitely many $p_i$ are non-zero.
Each application of the covariant derivative increases the
conformal weight $\lambda$ by one, and thus if
$\psi(x) \in \TT(\lambda, w)$, $\nabla^k \psi(x) \in \TT(\lambda + k, w)$.
Hence
$$ \eqalign{
-[L_m, \Phi^\bP(x)] &= \sum_k \psi(x)^{p_0} \ldots
p_k e^{mx} (\d + (\lambda + k) m + w)\nabla^k \psi(x)
 (\nabla^k \psi)^{p_k-1}(x) \ldots \cr
&= e^{mx} \bigg( \d \Phi^\bP(x)
+ (\lambda m + w) \sum_k p_k \Phi^\bP(x)
 + \sum_k k m p_k \Phi^\bP(x) \bigg) \cr
&= e^{mx} (\d + A(\bP) (\lambda m + w) + B(\bP) m) \Phi^\bP(x),
}	\eqno(2.13) $$
where $A(\bP) = \sum_k p_k$ and $B(\bP) = \sum_k k p_k$.
We conclude that the molecule
$\Phi^\bP(x) \in \TT(A(\bP) \lambda + B(\bP), A(\bP) w)$.

Continuing the physics terminology, $\nabla^k \psi$ can be thought of
as the {\it $k$:th shell} of the molecule and $p_k$ is the
corresponding occupation number. If we think of $\lambda$ as an
energy, we see that every particle in the $k$:th shell has the same
energy ($k + \lambda$), and hence the shells play the role of energy
levels. The {\it ground state} of a $p$-particle molecule is the state
with the lowest possible energy, i.e. the shells are filled from the
bottom up with totally $p$ particles. Other combinations are refered
to as excited states.

If $\psi$ is bosonic the occupation numbers can be arbitrary
non-negative integers.  The ground state is thus $\psi(x)^p \in
\TT(p\lambda, pw)$, and it is special because it does not involve the
gauge boson. We have thus constructed a map
$$ S^p \TT(\lambda, w) \longrightarrow \TT(p\lambda, pw),
	\eqno(2.14) $$
where $S^p$ denotes the $p$:th symmetric power.

The fermionic case is more interesting, because the occupation numbers
can only be zero or one. The $p$-fermion ground state $\Psi^{(p)}(x)$
is thus defined by $p_k = 1$ if $k<p$ and $p_k = 0$ if $k \ge p$.
Clearly,
$$ A(\bP) = \sum_{k=0}^{p-1} 1 = p
\qquad \hbox{and} \qquad
B(\bP) = \sum_{k=0}^{p-1} k = {p \choose 2} = {{p(p-1)} \over 2}.
	\eqno(2.15) $$
The ground state is again independent of the gauge bosons. This is not
as manifest as in the bosonic case, but it follows from the
anti-commutation relations.  For example, the two-fermion molecule
$$ \psi \nabla \psi = \psi (\d + \lambda \Gamma + w) \psi =
\psi \d \psi + (\lambda \Gamma + w) \psi^2,
	\eqno(2.16) $$
and the second term vanishes because $\psi^2 = 0$. An analogous
argument shows that
$$ \Psi^{(p)}(x) \equiv
\psi(x) \nabla \psi(x) \nabla^2 \psi(x) \ldots \nabla^{p-1} \psi(x)
= \psi(x) \d \psi(x) \d^2 \psi(x) \ldots \d^{p-1} \psi(x),
	\eqno(2.17) $$
which thus defines a map
$$ \Lambda^p \TT(\lambda, w) \longrightarrow
\TT(p\lambda + {p \choose 2}, pw),
	\eqno(2.18) $$
where $\Lambda^p$ denotes the $p$:th exterior power.

Feigin and Fuks\refto{13} call this map an operator in general
position. To see that our expression is the same as theirs requires
some extra work.  Up to a constant factor, (2.17) can be rewritten as
$$ \Delta(\d_1, \ldots, \d_p) (\psi(x_1) \ldots \psi(x_p)) \
\Big |_{x_1 = \ldots = x_p = x},
	\eqno(2.19) $$
where $\d_k$ is the derivative with respect to $x_k$ and the
Vandermonde determinant is
$$ \Delta(\d_1, \ldots, \d_p) = \prod_{1 \le i<j \le p} (\d_i - \d_j).
	\eqno(2.20) $$
We have e.g.
$$ (\d_1 - \d_2) \psi(x_1) \psi(x_2) \Big|_{x_1 = x_2 = x}
= \d(\psi(x) \psi(x)) - 2 \psi(x) \d \psi(x),
	\eqno(2.21) $$
and the first term vanishes because $\psi$ is a fermion.

For certain values of $\lambda$ and $w$ the size of the module can be
decreased by factoring out an invariant, which commutes with the
entire Witt algebra.  This is most easily explained in the Fourier
basis (2.6).  $\Theta_n$ is invariant if $\Theta \in \TT(\lambda = 1, w=n)$,
because then
$$ [L_m, \Theta_n] = (n - w) \Theta_{m+n} \equiv 0.
	\eqno(2.22) $$
Comparing with (2.18), we see that $\Psi^{(p)}_n$ is invariant provided that
$$ p\lambda + {{p(p-1)} \over 2} = 1,
\qquad \qquad pw = n.
	\eqno(2.23) $$
This equation selects infinitely many points on a curve in
$(\lambda, w)$-space,
which is defined by regarding $p$ as a continuous parameter.
For a given value of $\lambda$, (2.23) is a quadratic equation for $p$ which
generically has two complex solutions,
$$ p_\pm = -(\lambda - \half) \pm \sqrt{{(\lambda - \half)}^2 +2},
	\eqno(2.24) $$
We can ask if $(\lambda, w)$ lies on the intersection of two curves
of the type (2.23);
the answer is positive provided that there are two integers
$n_+$ and $n_-$ such that
$$ w = {n_+ \over p_+} = {n_- \over p_-}.
	\eqno(2.25) $$
This Diophantine equation has a solution if
$$ p_+ = \sqrt{ {{-2 n_-} \over {n_+}}},
\qquad
p_- = - \sqrt{ {{-2 n_+} \over {n_-}}},
	\eqno(2.26) $$
$(n_+ n_- < 0)$, or
$$ {(\lambda - \half)}^2 = -{{{(n_+ + n_-)}^2} \over {2 n_+ n_-}}.
	\eqno(2.27) $$

Let us now explain how this is connected to the discrete series of
Virasoro representations.  It terms of the fermion $\psi$ and its
conjugate $\psic$, which satisfy canonical anticommutation relations
(CAR),
$$ \eqalign{
\{\psic(x), \psi(y)\} &= \delta(x-y), \cr
\{\psic_m, \psi_n\} &= \delta_{m+n}, \cr
} \qquad \eqalign{
\{\psic(x), \psic(y)\} &= \{\psi(x), \psi(y)\} = 0. \cr
\{\psic_m, \psic_n\} &= \{\psi_m, \psi_n\} = 0. \cr
}	\eqno(2.28) $$
We can think of $\psi(x)$ as the anti-commuting coordinates of some
vector space $V_\psi$ and of $\psic(x)$ as the corresponding
derivative. The envelopping algebra of (2.28) is then the algebra of
all differential operators on this space, i.e. a fermionic Weyl
algebra.  $Vect(1)$ acts on this Weyl algebra as the vector field
$$ L_m = \int dx \; \e^{mx} \; \psic(x) (\d + \lambda m + w) \psi(x)
= \sum_s (-s + \lambda m + w) \psic_{m-s} \psi_s.
	\eqno(2.29) $$
This means that we have a representation of the Witt algebra on the
fermionic Weyl algebra. If the Witt algebra were finite-dimensional,
it would inherit a representation for any representation of (2.28),
but this is not quite true in the infinite-dimensional setting.

We have on purpose not specified the domain of the summation variable
$s$.  Since $m$ is an integer it is clear that $s$ must run over
numbers which differ by integers, i.e. $s \in \Ints + \alpha$ for some
real number $\alpha$.  It is natural to demand that $\alpha = 1/2$,
because this choice makes $\psi$ into a kind of ``one-dimensional
spinor''. Moreover, for $\alpha$ integer or half-integer $\psi$ and
$\psic$ appear symmetrically.  However, the ultimate reason for this
choice is that it yields the correct Kac table below.

The irreducible representation of the CAR is the {\it fermionic Fock
module}, which is obtained from the envelopping algebra of (2.28) by
deleting all strings containing elements of negative degree.
Equivalently, we introduce the {\it vacuum} $\vac$, satisfying
$$ \psi_{-s} \vac = \psic_{-s} \vac = 0, \qquad \qquad \forall s > 0.
	\eqno(2.30) $$
We also introduce the shifted vacuum $\qvac$, defined by
$$ \qvac = \psi_{q-1/2} \ldots \psi_{1/2} \vac,
	\eqno(2.31) $$
which satisfies
$$
\psi_{q-s} \vac = \psic_{-q-s} \vac = 0, \qquad \qquad \forall s > 0.
	\eqno(2.32) $$
$q$ can also be negative; $\psi$ is then replaced by $\psic$ in (2.31).

The inherited $Vect(1)$ module is defined by
$$ \eqalign{
L_m \qvac &=
\sum_{s=q+1/2}^{m+q-1/2} (-s + \lambda m + w) \psic_{m-s} \psi_s \qvac \cr
L_0 \qvac &= h \qvac \cr
L_{-m} \qvac &= 0 ,
}	\eqno(2.33) $$
together with
$$ \eqalign{
[L_m, \psi_n] &= (n + (1-\lambda) m - w) \psi_{m+n}, \cr
[L_m, \psic_n] &= (n + \lambda m + w) \psic_{m+n}. \cr
}	\eqno(2.34) $$
Because of the infinite dimensionality, (2.33) and (2.34) are actually
not representations of the Witt algebra but of the Virasoro algebra
(1.4), because our procedure is nothing but the standard normal
ordering recipe.  By demanding that $L_0$ should have a finite
eigenvalue, we have subtracted off an infinite constant.  After a
straightforward calculation it is found that
$$ \eqalign{
c &= -2 (6 \lambda^2 - 6 \lambda + 1) = 1 - 12(\lambda - \half)^2 \cr
2h &= (w - q)^2 - (\lambda - \half)^2
}	\eqno(2.35) $$

The Fock module decomposes into modules where the difference between
the numbers of $\psi$'s and $\psic$'s is fixed, because the fermionic
number operator commutes with $L_m$. In a obvious notation we denote
by $\psi^q \vac$ the homogeneous component where this difference
equals $q$; $\qvac$ is the vacuum in this submodule.

A {\it singular vector} is an element which satisfies the same
conditions as the vacuum. Given the invariant $\Psi^{(p)}_t$ above, we
obtain the singular vector by applying it to the vacuum. $\Psi^{(p)}_t
\vac$ is hence a singular vector in the module $\psi^p \vac$, and
${(\Psi^{(p)}_t)}^j \vac$ is singular in $\psi^{jp} \vac$. Explicitly
$$ L_{-m} {(\Psi^{(p)}_t)}^j \vac =
[L_{-m}, {(\Psi^{(p)}_t)}^j] \vac + {(\Psi^{(p)}_t)}^j L_{-m} \vac = 0 + 0.
	\eqno(2.36) $$
If $p$ is odd, $\Psi^{(p)}_t$ is fermionic and the singular vector vanishes
identically. The interesting case is thus $p$ even, and hence the procedure
can be continued to $j$ half-integer.
With $j = s/2$, $s \in \Naturals$, there is a singular vector in
$\psi^{sp/2} \vac$, which is characterized by (2.23) and (2.35), where
$q = sp/2$. Hence, the module is reducible if
$$ \eqalign{
c &= 1 - 12 {\Big({1\over p} - {p \over 2} \Big)}^2, \cr
2h &= {\Big( {t \over p} - {{sp} \over 2} \Big)}^2
- {\Big( {1\over p} - {p \over 2} \Big)}^2.
}	\eqno(2.37) $$
By eliminating $p$ we obtain Kac'
formula for reducibility of the Verma module. In particular, for the
the special values of $\lambda$ given by (2.27),
$$ \eqalign{
c &= 1 - {{6{(m - n)}^2} \over {mn}} \cr
h &= {{{(tm-sn)}^2 - {(m-n)}^2} \over {4mn}} \cr
}	\eqno(2.38) $$
where $m = n_+$, $n = - n_-$, $m, n, s, t\in \Naturals$. This is the
discrete series of irreducible Virasoro representations.\refto{2}
Eq. (2.38) was obtained from (2.37) by substituting $p = p_+$ from
(2.26). If we instead use $p = p_-$, we find the same formula for $h$
but with $t$ and $s$ interchanged. This is a consistency check on the
formula.

Our construction can not really be considered as a proof of Kac'
formula, because there is no guarantee that the singular vectors are
non-trivial. There are two possibilities to find singular vectors
which are not covered by the list (2.38).  First we can use fermions
whose Fourier components are not half-integers.  Second, we could
consider the singular vectors ${\big( \Psi^{(p)}_t \big)}^{s/k} \vac$
for every $p$ divisible by $k$. Since these generalizations would lead
outside Kac' formula, we conclude that these new singular vectors are
trivial.

We can consider more complicated Fock modules by including the gauge
boson $\Gamma(x)$ and its canonical conjugate $\Gammac(x)$. Their
Fourier components must be labelled by integers, and they satisfy
$$
[\Gamma_m, \Gammac_n] = \delta_{m+n},
\qquad
[\Gamma_m, \Gamma_n] =[\Gammac_m, \Gammac_n] = 0.
	\eqno(2.39) $$
The contribution to the $Vect(1)$ generators is
$$ L_m
= \sum_{n \in \Ints} (-n+m) \Gammac_{m-n} \Gamma_n - m^2 \Gammac_m,
	\eqno(2.40) $$
and the vacuum satisfies
$$ \eqalign{
\Gamma_{-n} \vac &= \Gammac_{-n} \vac = 0,
\qquad \qquad \forall n > 0. \cr
\Gammac_0 \vac &= 0.
}	\eqno(2.41) $$
When (2.40) is applied to the vacuum we find the contribution $c = 2$,
$h = 0$, which should be added to (2.35). In the combined Fock space
containing both fermions and gauge bosons, new singular vectors can be
constructed by demanding that some of the excited states (2.12) be
invariant and applying them to the vacuum. The relevance of these more
complicated modules is not clear to us.

Shells can also be defined by using bosons rather than fermions, but
this is less interesting. In one dimension bosons and fermions are
equivalent,\refto{14} wherefore the Feigin-Fuks construction can be
rewritten in terms of bosons; they are essentially the vertex
operators.\refto{15} The fermion approach seems nicer because it
generalizes to higher dimensions. Moreover, if we seriously believe
that $Vect(N)$ is relevant to physics, it is natural to expect that
the interesting representations deal with fundamental fermions and
gauge bosons, because that is the kind of fundamental particles which
are seen experimentally.

\endchapter

\header{3. Kac-Moody algebras}

The construction of the previous section immediately suggests a
generalization to Kac-Moody algebras. Recall that the Kac-Moody
algebra is a central extension of the loop algebra based on a
finite-dimensional Lie group $\oj$.\refto{16} The brackets are
$$ [J^a_m, J^b_n] = f^{abc} J^c_{m+n} + k m \delta^{ab} \delta_{m+n}
	\eqno(3.1) $$
where $f^{abc}$ are the totally skew-symmetric structure constants of
$\oj$ and $\delta^{ab}$ is the Killing metric. Because of this metric,
there is no need to distinguish between upper and lower $\oj$ indices.

Actually, we are only interested in representations which admit an
intertwining action of $Vect(1)$, because the generators must
transform covariantly under arbitrary reparametrizations of the loop.
The interesting algebra is thus the semi-direct product
$Vect(1) \IX Map(1, \oj)$,
$$ [L_m, J^b_n] = n J^b_{m+n},
	\eqno(3.2) $$
and its central extensions. The classical representations of (3.1--2) are
$$ [J^a_m, \psi^\alpha_n] = - M^{\alpha a}_\beta \psi^\beta_{m+n}
	\eqno(3.3) $$
where $M^a = (M^{\alpha a}_\beta)$ are matrices of some
finite-dimensional $\oj$ representation, and $\psi^\alpha_n$
transforms as (2.6) under $Vect(1)$. Alternatively, the transformation
law (3.3) can be recast in real space
$$ [J^a_m, \psi(x)] = - \emx M^a \psi(x),
	\eqno(3.4) $$
with representation indices suppressed. We denote this representation
by $\TT(\lambda, w) \times M$.

To construct a first-order differential operator we must covariantize
the map (2.8) with respect to $\oj$. It is straight-forward to check
that the {\it gauge connection} $\omega^a(x)$, which transforms as
$$ [J^a_m, \omega^b(x)] = \emx ( f^{abc} \omega^c(x) + m \delta^{ab})
	\eqno(3.5) $$
and as the $Vect(1)$ primary field $\TT(1,0)$, is defined consistently.
The covariant derivative is now the map
$$ \eqalign{
\TT(\lambda, w) \times M &\longrightarrow \TT(\lambda+1, w) \times M \cr
D \psi(x) &= \nabla \psi(x) + \omega^a(x) M^a \psi(x)
}	\eqno(3.6) $$
where $\nabla$ is given by (2.8).

A molecule can be constructed from fundamental particles transforming in
the $r$-dimensional representation $M$ in the following fashion.
$$ \Phi^\bP(x) = {\psi^1(x)}^{p^1_0} \ldots {\psi^r(x)}^{p^r_0}
{(D \psi^1(x))}^{p^1_1} \ldots {(D \psi^r(x))}^{p^r_1}
{(D \psi^1(x))}^{p^1_2} \ldots {(D \psi^r(x))}^{p^r_2} \ldots,
	\eqno(3.7) $$
where $\bP = \{p^\alpha_k: 1 \le \alpha \le r, k \ge 0 \}$, and only
finitely many $p^\alpha_k$ are non-zero.  Because there are now $r$
different states with the same energy, we say that all of them belong
to the same shell, and $p_k = \sum_{\alpha=1}^r p^\alpha_k$ is the
total occupation number of the $k$:th shell.  The ground states are
again independent of both the connection $\Gamma(x)$ and the gauge
connection $\omega^a(x)$. In the bosonic case we thus have a map
$$ S^p (\TT(\lambda, w) \times M) \longrightarrow
\TT(p\lambda, pw) \times S^p M,
	\eqno(3.8) $$
and by appropriate symmetrization of the $\oj$ indices, the image of
this map can be restricted to any representation $N \subset S^p M$.

The case of fermions is again more interesting. The occupation numbers
can now range between zero (empty shell) and $r$ (full shell).  A
ground state, which we denote by $\Psi^{(p,b)}(x)$, consists of $pr+b$
fermions, $0 \le b < r$, where $p$ is the number of full shells and
$b$ is the number of fermions in the valence shell. It is clear that a
full shell, e.g.
$$ \psi^1(x) \psi^2(x) \ldots \psi^r(x)
	\eqno(3.9) $$
is a gauge singlet, so full shells do not contribute to the gauge
representation. Because the gauge bosons factor out in the ground state,
we have constructed a map
$$ \Lambda^{pr+b} (\TT(\lambda, w) \times M) \longrightarrow
\TT((pr+b)\lambda + r {p \choose 2} + b p, (pr+b)w) \times \Lambda^b M.
	\eqno(3.10) $$
By appropriate symmetrization of the $\oj$ indices, the image of this
map can
be restricted to any $\oj$ representation $N \subset \Lambda^b M$.

An invariant must be a $\oj$ singlet, which is possible only if $b=0$.
$\Psi^{(p,0)}_n$ is an invariant in
$\Lambda^{pr} (\TT(\lambda, w) \times M)$
provided that
$$
pr\lambda + r {{p(p-1)} \over 2} = 1,
\qquad prw = n,
	\eqno(3.11) $$
for some integer $n$. The discrete series of $\lambda$'s which admit
two invariants
are again characterized by $n_+ p_- = n_- p_+$, i.e.
$$ p_+ = \sqrt{ {{-2 n_-} \over {r n_+}}},
\qquad
p_- = - \sqrt{ {{-2 n_+} \over {r n_-}}},
	\eqno(3.12) $$
and
$$ {(\lambda - \half)}^2 = -{{{(n_+ + n_-)}^2} \over {2 r n_+ n_-}},
	\eqno(3.13) $$
where $n_+ n_- < 0$. The only difference compared to (2.27) is
the factor $r$ in
the denominator.

This result can immediately be used to find singular vectors in
fermionic Fock modules. Introduce the canonical conjugate of the
fermions,
$$
\{\psic_{\beta m}, \psi^\alpha_n\} = \delta^\alpha_\beta \delta_{m+n},
\qquad
\{\psic_{\alpha m}, \psic_{\beta n} \} =
\{\psi^\alpha_m, \psi^\beta_n\} = 0,
	\eqno(3.14) $$
and let $\vac$ be the vacuum as in Sec. 2. Other vacua are constructed
from this vector by filling the states immediately above the Fermi
level. Since there is room for $r$ fermions with the same momentum,
the following expression is a gauge singlet
$$ \qvac = \psi^1_{q-1/2} \ldots \psi^r_{q-1/2} \ldots
\psi^1_{1/2} \ldots \psi^r_{1/2} \vac.
	\eqno(3.15) $$
In the notation of Sec. 2, this is the vacuum of the module $\psi^{qr}
\vac$.  Other vacua carrying non-trivial representations of $\oj$ can
of course also be constructed, but we limit ourselves to the class
(3.15).

It can be checked that
the following generators satisfy Virasoro and Kac-Moody algebras,
respectively.
$$ \eqalign{
L_m \qvac &= \sum_{s = q+1/2}^{m+q-1/2}
(-s + \lambda m + w) \psic_{\alpha,m-s} \psi^\alpha_s \qvac \cr
L_0 \qvac &= h \qvac \cr
L_{-m} \qvac &= 0 \cr
{}\cr
J^a_m \qvac &=
\sum_{s = q+1/2}^{m+q-1/2} M^{\alpha a}_\beta
\psic_{\alpha,m-s} \psi^\beta_s \qvac \cr
J^a_0 \qvac =
J^a_{-m} \qvac &= 0. \cr
{}
}	\eqno(3.16) $$

The parameters are given by
$$ \eqalign{
c &= r \Big( 1 - 12(\lambda - \half)^2 \Big) \cr
k &= Q_M \cr
2h &= r \Big( (w - q)^2 - (\lambda - \half)^2 \Big) \cr
}	\eqno(3.17) $$
where $Q_M$ is the value of the quadratic Casimir in the
representation $M$, i.e. $\tr M^a M^b = Q_M \delta^{ab}$.

The singular vector in $\psi^{jpr} \vac$, $p$ even, is obtained by
applying the invariant molecule with $p$ full shells $j$ times
to the vacuum, i.e.
$ {(\Psi^{(p,0)}_t)}^j \vac $. This singular vector lies
in the module with vacuum $\state{jp}$. We can formally continue
this to the case that $2 j \in \Ints$. If we set $j = s/2$,
$q = sp/2$, the reducible modules are
$$ \eqalign{
c &= r \Big( 1 - 12{\big({1 \over {rp}} - {p \over 2} \big)}^2 \Big) \cr
k &= Q_M \cr
2h &= r \Big( {\big( {t \over {pr}} - {{sp} \over 2} \big)}^2
- {\big( {1 \over {rp}} - {p \over 2} \big)}^2 \Big) \cr
}	\eqno(3.18) $$

In particular in the case that there are two different singular vectors for
the same value of $c$,
$$ \eqalign{
c &= r - {{6{(m - n)}^2} \over {mn}} \cr
k &= Q_M \cr
h &= {{{(tm-sn)}^2 - {(m-n)}^2} \over {4mn}}, \cr
}	\eqno(3.19) $$
where $m = n_+$ and $n = -n_-$. The only difference compared to the
previous section, apart from the appearence of the Kac-Moody algebra,
is that the central charge has been increased by $r-1$.

Just as in the previous section, we can not be sure that the singular
vectors constructed above are non-trivial. Moreover, other
possibilities arise in a module whose vacuum is not a gauge singlet.
The purpose of this section is thus not to present an exhaustive list
of all analogs of Kac' table for every Kac-Moody algebra, but rather
to illustrate that molecules yield a simple method for concretely
constructing singular vectors in quite general situations. This will
be further emphasized in the next section, where we turn to higher
dimensions.

\endchapter

\header{4. Invariants in several dimensions}

We now generalize the concept of a molecule to more than one dimension.
In a plane-wave basis, the generators of $Vect(N)$ are
$$ L^\mu(m) = \emx \d^\mu,
	\eqno(4.1) $$
where $x = (x_1, \ldots, x_N)$, $[\d^\mu, x_\nu] = \delta^\mu_\nu$, and
$m = (m^1, \ldots, m^N)$ belongs to an $N$-dimensional
imaginary lattice $\Lambda$.
The algebra thus reads
$$ [L^\mu(m), L^\nu(n)] = n^\mu L^\nu(m+n) - m^\nu L^\mu(m+n).
	\eqno(4.2) $$
Globally, this is the algebra of vector fields on an $N$-dimensional
torus, but since any two manifolds of the same dimension are locally
diffeomorphic, (4.2) applies locally to any $N$-dimensional manifold.

An important class of $Vect(N)$ representations are {\it tensor fields}
(or densities), which are
constructed from $gl(N)$ representations as follows.
Assume that $\{T^\mu_\nu\}_{\mu,\nu=1}^N$ satisfies $gl(N)$, i.e.
$$ [T^\mu_\sigma, T^\nu_\tau] =
 \delta^\mu_\tau T^\nu_\sigma - \delta^\nu_\sigma T^\mu_\tau.
	\eqno(4.3) $$
Then it is easy to check that
$$ L^\mu(m) = \emx \Big(\d^\mu + w^\mu + m^\sigma T^\mu_\sigma \Big)
	\eqno(4.4) $$
satisfies $Vect(N)$, where $w^\mu$ is a constant vector which is
defined modulo $\Lambda$. Hence we have a $Vect(N)$ representation for
each $gl(N)$ representation.  From a $gl(N)$ tensor with $p$ upper and
$q$ lower indices and conformal weight $\lambda$, we obtain the
$Vect(N)$ module $\TT^p_q(\lambda, w)$ with action
$$ \eqalign {
[L^\mu(m), \psi^\nup_\tauq(x)] &=
-\emx \Big( (\d^\mu + w^\mu + \lambda m^\mu) \psi^\nup_\tauq(x) \cr
&+ \sum_{i=1}^p m^{\nu_i} \psi^{\argsx\nu p \mu}_\tauq(x)
- \sum_{j=1}^q \delta^\mu_{\tau_j} m^\sigma
\psi^\nup_{\argsx\tau q\sigma}(x) \Big).
}	\eqno(4.5) $$
We write $\TT^p_q(\lambda) = \TT^p_q(\lambda, 0)$. This equation
clearly reduces to (2.4) in one dimension, with $\lambda$ replaced by
$\lambda + p - q$.  In the Fourier basis,
$$ \eqalign {
[L^\mu(m), \psi^\nup_\tauq(n)] &=
(n^\mu - w^\mu + (1-\lambda) m^\mu) \psi^\nup_\tauq(m+n) \cr
&- \sum_{i=1}^p m^{\nu_i} \psi^{\argsx\nu p \mu}_\tauq(m+n)
+ \sum_{j=1}^q \delta^\mu_{\tau_j} m^\sigma
\psi^\nup_{\argsx\tau q\sigma}(m+n).
}	\eqno(4.6) $$
In particular, we have for a scalar field
$$ [L^\mu(m), \psi(n)] = (n^\mu - w^\mu + (1-\lambda) m^\mu) \psi(m+n).
	\eqno(4.7) $$
The adjoint is $\TT^1_0(1)$.

It is sometimes convenient to describe $Vect(N)$ more invariantly. Define
$$ L(f_\mu) = \sum_{m \in \Lambda} f_\mu(m) L^\mu(m)
	\eqno(4.8) $$
for each function $f_\mu(x)$ with Fourier coefficients $f_\mu(m)$.
The algebra then takes the form
$$ [L(f_\mu), L(g_\mu)] = L( f_\nu \d^\nu g_\mu - g_\nu \d^\nu  f_\mu)
= L(f_\nu \d^\nu g_\mu) - L(g_\nu \d^\nu f_\mu).
	\eqno(4.9) $$
Tensor fields are given by
$$ L(f_\mu) = f_\mu(x) \d^\mu + f_\mu(x) w^\mu(x)
+ \d^\sigma f_\mu(x) T^\mu_\sigma,
	\eqno(4.10) $$
where the function $w^\mu(x)$ satisfies
$$ \d^\mu w^\nu(x) - \d^\nu w^\mu(x) + [w^\mu(x), w^\nu(x)] = 0.
	\eqno(4.11) $$
This zero-curvature condition has the solution
$w^\mu(x) = u^{-1}(x) \d^\mu u(x)$.
Substituting this solution into (4.10), we obtain
$$ L(f_\mu) =  u^{-1}(x) \Big( f_\mu(x) \d^\mu +
\d^\sigma f_\mu(x)  T^\mu_\sigma \Big) u(x)
	\eqno(4.12) $$
On the torus we can make the non-trivial choice $u(x) = \exp(w \cdot x)$,
which gives back (4.4). An advantage of the formulation (4.12) is
that it suggests a generalization of $w$ when $f_\mu(x)$ is expanded
in another set of basis functions.  This is important because Kac'
formula (2.37) is essentially a relation between the parameters
$\lambda$ and $w$.  E.g., in three dimensions, $f_\mu(x) =
f_\mu(r,\theta,\varphi)$ can be expanded in the basis $r^n
Y_{lm}(\theta, \varphi)$, $n \in \Ints$, and a non-trivial function is
$u(x) = r^w$, $0 < w < 1$. For the rest of this section we limit
ourselves to the torus.

The {\it connection} is a central extension of the representation
$\TT^2_1(0)$, with transformation law
$$ \eqalign{
[L^\mu(m), \Gamma^{\sigma\nu}_\tau(x)] &=
\emx \Big( -\d^\mu \Gamma^{\sigma\nu}_\tau(x)
- m^\sigma \Gamma^{\mu\nu}_\tau(x) \cr
&- m^\nu \Gamma^{\sigma\mu}_\tau(x)
+ \delta^\mu_\tau m^\rho \Gamma^{\sigma\nu}_\rho(x)
+ \delta^\mu_\tau m^\sigma m^\nu \Big),
}	\eqno(4.13) $$
and the {\it covariant derivative} is the map
$$ \eqalign{
\TT^p_q(\lambda, w) &\longrightarrow \TT^{p+1}_q(\lambda, w) \cr
\nabla^\nu \psi(x) &= (\d^\nu + w^\nu) \psi(x)
+ \Gamma^{\sigma\nu}_\tau(x) [T^\tau_\sigma, \psi(x)]
}	\eqno(4.14) $$
For a scalar field,
$$ \nabla^\nu \psi(x)
= (\d^\nu + w^\nu + \lambda \Gamma^{\sigma\nu}_\sigma(x)) \psi(x).
	\eqno(4.15) $$

Just as in Sec. 2 we can now build a molecule out of fundamental particles
and their covariant derivatives at the same point $x$.
For simplicity we limit ourselves to scalar fields and fermions.
A fermion in the $k$:th shell enters through the expression
$$ \nabla^{\nu_1} \ldots \nabla^{\nu_k} \psi \in \TT^k_0(k \lambda, k w)
	\eqno(4.16) $$
which depends on the connection. The ground state with the $p$
lowest shells being filled is
$$ \eqalign{
\Psi^{(p)}(x)
&= \psi(x) \nabla^{\nu_1} \psi(x) \ldots \nabla^{\nu_N} \psi(x)
\nabla^{\sigma_{11}} \nabla^{\tau_{11}} \psi(x)
\nabla^{\sigma_{12}} \nabla^{\tau_{12}} \psi(x) \ldots \cr
&= \psi(x) \prod_{i=1}^N \nabla^{\nu_i} \psi(x)
\prod_{j=1}^N \prod_{k=1}^j \nabla^{\sigma_{jk}}
\nabla^{\tau_{jk}} \psi(x) \ldots. \cr
}	\eqno(4.17) $$

In this ground state all references to the connection vanishes due to
anti-symmetry, and therefore we can replace covariant derivatives with
usual ones. The expression (4.16) is replaced by
$$ \d^{\nu_1} \ldots \d^{\nu_k} \psi,
	\eqno(4.18) $$
which is manifestly symmetric in $\nu_1 \ldots \nu_k$,
and the ground state becomes
$$ \Psi^{(p)}(x)
= \psi(x) \prod_{i=1}^N \d^{\nu_i} \psi(x)
\prod_{j=1}^N \prod_{k=1}^j
\d^{\sigma_{jk}} \d^{\tau_{jk}} \psi(x) \ldots.
	\eqno(4.19) $$
The number of different states in the $k$:th shell is thus equal to
the number of symmetric combinations of $k$ indices which can take $N$
different values, i.e.
$$ n_k = {{N-1+k} \choose k} = {{N-1+k} \choose {N-1}}.
	\eqno(4.20) $$
In the molecule with $p$ full shells, (4.20) equals the occupation
number for the $p$ lowest shells. Each fermion in the $k$:th shell
contributes with $k$ upper indices, wherefore
$$ \Psi^{(p)}(x) \in \TT^{B_N(p)}_0(A_N(p) \lambda, A_N(p) w),
	\eqno(4.21) $$
where
$$ \eqalign{
A_N(p) &= \sum _{k=0}^{p-1} n_k
= \sum _{k=0}^{p-1} {{N-1+k} \choose {N-1}}
= {{N-1+p} \choose N}, \cr
{}\cr
B_N(p) &= \sum _{k=0}^{p-1} k n_k
= \sum _{k=0}^{p-1} k {{N-1+k} \choose {N-1}} =
N \sum _{j=0}^{p-2} {{N+j} \choose N}
= N {{N-1+p} \choose {N+1}}. \cr
}	\eqno(4.22) $$
Since the total number of fermions in the $p$-shell molecule
is $A_N(p)$, we have constructed a map
$$ \Lambda^{A_N(p)} \TT^0_0(\lambda, w) \longrightarrow
\TT^{B_N(p)}_0(A_N(p) \lambda, A_N(p) w).
	\eqno(4.23) $$

The range of this map is actually a submodule, characterized by certain
symmetries. E.g., the three-shell ground state in two dimensions,
$$
\Psi^{\nu_1 \nu_2 \sigma_{11} \tau_{11} \sigma_{12}
\tau_{12} \sigma_{22} \tau_{22}}
= \, \psi \, \d^{\nu_1} \psi \, \d^{\nu_2} \psi
\, \d^{\sigma_{11}} \d^{\tau_{11}} \psi
\, \d^{\sigma_{12}} \d^{\tau_{12}} \psi
\, \d^{\sigma_{22}} \d^{\tau_{22}} \psi.
	\eqno(4.24) $$
is skew in $\nu_1$ and $\nu_2$, symmetric in $\sigma_{ij}$
and $\tau_{ij}$, and
skew under interchange of any pairs
$ \sigma_{ij} \tau_{ij} \xxx \sigma_{kl} \tau_{kl}$.

Some more work is needed before (4.19) can be used to construct an
invariant.  Consider the submodule $\Omega_p \subset \TT^0_p(1)$
consisting of totally skew tensors with $p$ lower indices. There is a
sequence of maps
$$ \Omega_N \mapright{\bar d_N} \Omega_{N-1}
\mapright{\bar d_{N-1}} \ldots
\mapright{\bar d_1} \Omega_0 \mapright{\bar d_0} \Complexs,
	\eqno(4.25) $$
given by
$$ \eqalign{
(\bar d_p \phi(x))_{\nu_1\ldots\nu_{p-1}}
&= \d^\sigma \phi(x)_{\nu_1\ldots\nu_{p-1}\sigma}
\qquad (p \ge 1) \cr
\bar d_0 \phi = \phi(n = 0).
}	\eqno(4.26) $$
Moreover, $\bar d_{p-1} \bar d_p = 0$.
The elements in $\Omega_p$ can be thought of as $p$-dimensional
integral operators,
$$ \phi_{\nu_1 \ldots \nu_p}(n) =
\underbrace{\int\ldots\int}_p dx_{\nu_1} \ldots dx_{\nu_p} \; \enx,
	\eqno(4.27) $$
and $\bar d_p$ is the boundary map. There is also a dual sequence
of differential forms.

The important map for our purposes is the last one. In the Fourier basis,
an element in $\Omega_0 = \TT^0_0(1)$ satisfies
$$ [L^\mu(m), \Theta(n)] = n^\mu \Theta(m+n) ,
	\eqno(4.28) $$
and it is clear that $\Theta(0)$ is an invariant; it transforms trivially
and therefore we can consistently set it equal to zero.
This can be slightly generalized by a shift in $n$.
$$ [L^\mu(m), \Theta(n)] = (n^\mu - w^\mu) \Theta(m+n) ,
	\eqno(4.29) $$
which shows that $\Theta(n)$ is an invariant in $\TT^0_0(1, n)$.

The object (4.19), which only has upper indices with certain
symmetries, can be transformed into a scalar provided that we
introduce a field with lower indices.  The resulting invariant depends
on this new field, which is undesirable unless it can be introduced in
a canonical manner. However, there is a unique object with $N$ skew
lower indices, namely the permutation symbol $\epsilon_{\nu_1 \ldots \nu_N}$,
which is defined by $\epsilon_{1 2 \ldots N} = 1$ and total
skewness. The permutation symbol can be considered as a constant
tensor field
$\epsilon_{\nu_1 \ldots \nu_N}(x) \in \Omega_N \subset \TT^0_N(1)$.
To see this, consider
$$ [L^\mu(m), \epsilon_{\nu_1 \ldots \nu_N}(x)] =
-\emx \Big( (\d^\mu + m^\mu) \epsilon_{\nu_1 \ldots \nu_N}(x)
- \sum_{i=1}^N \delta^\mu_{\nu_i} m^\sigma
\epsilon_{\nu_1 \ldots \sigma \ldots \nu_N}(x) \Big).
	\eqno(4.30) $$
If $\mu = \nu_1$, say, the only term that survives in the last sum is
$m^\sigma \epsilon_{\sigma \nu_2 \ldots \nu_N}$,
which cancels the first term because the permutation symbol is
non-zero only when $\sigma = \nu_1$.  In one dimension the permutation
symbol becomes the one-component vector
$\epsilon_\nu = \epsilon_1 = 1$, transforming as
$$ [L_m, \epsilon] = -\emx ( (\d + m) \epsilon - m \epsilon ) = 0.
	\eqno(4.31) $$
Dually, we may regard the permutation symbol as a tensor field with
upper indices and weight $-1$;
$\epsilon^{\nu_1 \ldots \nu_N}(x) \in \TT^N_0(-1)$.

The desired scalar field is now formed by contracting (4.19)
with the permutation symbol in a way which respects the symmetries.
{}From the two-dimensional three-shell ground state (4.24), we obtain
$$
\epsilon_{\nu_1 \nu_2}
\, \epsilon_{\sigma_{11} \sigma_{12}}
\, \epsilon_{\tau_{11} \sigma_{22}}
\, \epsilon_{\tau_{12} \tau_{22}}
\, \psi \, \d^{\nu_1} \psi \, \d^{\nu_2} \psi
\, \d^{\sigma_{11}} \d^{\tau_{11}} \psi
\, \d^{\sigma_{12}} \d^{\tau_{12}} \psi
\, \d^{\sigma_{22}} \d^{\tau_{22}} \psi.
	\eqno(4.32) $$
A little thought reveals that this procedure is well-defined for arbitrary
molecules with full shells.
There are $B_N(p)$ upper indices to contract. Since each
$\epsilon_{\nu_1 \ldots \nu_N}$
has $N$ lower indices, we need a total of $B_N(p)/N$ permutation
symbols, each contributing unity to the parameter $\lambda$ and
nothing to $w$. Hence we have a map
$$ \Lambda^{A_N(p)} \TT^0_0(\lambda, w) \longrightarrow
\TT^0_0(A_N(p) \lambda + {1 \over N} B_N(p), A_N(p) w).
	\eqno(4.33) $$

The map (4.33) defines an invariant provided that
$$
A_N(p) \lambda + {1 \over N} B_N(p) = 1,
\qquad\qquad
A_N(p) w^\mu = n^\mu,
	\eqno(4.34) $$
for some $n \in \Lambda$. Equivalently,
$$
\lambda - {1 \over {N+1}} = {1 \over {A_N(p)}} - {p \over {N+1}},
\qquad\qquad
w^\mu = {{n^\mu} \over {A_N(p)}},
	\eqno(4.35) $$
because $B_N(p) = N (p-1) A_N(p) / (N+1)$. Note that every solution to
the second equation is parallel to $w^\mu$.  If we introduce $\kappa =
(N+1) \lambda - 1$, the first equation takes the form
$$ (p + \kappa) {{N+p-1} \choose N} = N+1,
	\eqno(4.36) $$
i.e.
$$ p (p+1) \ldots (p+N-1) (p + \kappa) = (N+1)!.
	\eqno(4.37) $$
This is a polynomial equation of degree $N+1$, which generically has
$N+1$ complex solutions $p_i$. The maximal number of invariants in
this module is thus $N+1$, which is reached if all $p_i$ are real and
different and satisfy the simultaneous Diophantine equations
$$ w^\mu \propto {{n_i} \over {A_N(p_i)}} = {{n_j} \over {A_N(p_j)}}
\qquad i, j = 1, \ldots N+1.
	\eqno(4.38) $$

Instead of solving these solutions, we will at once generalize to the
case that we have an action of the the {\it gauge algebra} $Map(N, \oj)$.
As in Sec. 3, there are additional brackets
$$ \eqalign{
[J^a(m), J^b(n)] &= f^{abc} J^c(m+n), \cr
[L^\mu(m), J^b(n)] &= n^\mu J^b(m+n).
}	\eqno(4.39) $$
If the field $\psi(x)$ takes values in an $r$-dimensional
representation of $\oj$, the entire analysis above is unchanged except
that the RHS of (4.37) is replaced by $(N+1)!/r$, because each shell
can be filled with $r$ times as many fermions as before.

For $N=1$, $p_i = n_i x$ and (4.37) reads
$$ n_1 n_2 x^2 = -{2 \over r},
\qquad\qquad
(n_1 + n_2)x = -\kappa
	\eqno(4.40) $$
with the solution
$$ n_1 n_2 < 0,
\qquad
x = \sqrt{-{ 2 \over {r n_1 n_2}}},
\qquad
\kappa^2 = -{{2 {(n_1 + n_2)}^2} \over {r n_1 n_2}}.
	\eqno(4.41) $$
This is the discrete series found in (3.13).

For $N=2$,
$$ \eqalign{
1 + \kappa &= -(p_1 + p_2 + p_3), \cr
\kappa &= p_1 p_2 + p_1 p_3 + p_2 p_3, \cr
{6 \over r} &= p_1 p_2 p_3,
}	\eqno(4.42) $$
where the $p_i$ are related by
$$ {{p_i (p_i+1)} \over 2} = n_i x,
\qquad n_i \in \Ints.
	\eqno(4.43) $$

As a last example, the specialization of (4.37) to $N=3$ reads
$$ \eqalign{
3 + \kappa &= - \sum_{i=1}^4 p_i, \cr
2 + 3\kappa &= \sum_{1\le i < j \le4} p_i p_j \cr
2 \kappa &= - \sum_{1\le i < j < k \le4} p_i p_j p_k \cr
{24 \over r} &= - p_1 p_2 p_3 p_4,
}	\eqno(4.44) $$
where the $p_i$ are related by
$$ {{p_i (p_i+1) (p_i+2)} \over 6} = n_i x,
\qquad n_i \in \Ints.
	\eqno(4.45) $$

It is not clear to us that any solution exists when $N \ge 2$, but if
it does, it generalizes the discrete series of $\lambda$'s. Even if it
is impossible to find the maximal number of invariants, we can give up
a few of these equations and still obtain a discrete set of
exceptionally small modules, by factoring out the remaining
invariants.

The shell construction can be generalized in several directions.
Instead of starting with a scalar field, we could use a vector field
$\psi^\nu(x)$. Because $\nu$ can take $N$ different values, each shell
can be occupied by $N$ times as many fermions as in the scalar case.
Explicitly, the $p$-shell molecule has the form
$$
\prod_{i=1}^N \psi^{\nu_i}
\prod_{j=1}^N \prod_{k=1}^N \nabla^{\sigma_{j}} \psi^{\tau_{k}} \ldots
= \prod_{i=1}^N \psi^{\nu_i}
\prod_{j=1}^N \prod_{k=1}^N \d^{\sigma_{j}} \psi^{\tau_{k}} \ldots.
	\eqno(4.46) $$
The molecule can be contracted with the permutation symbol to obtain an
invariant.
Since it now takes $N A_N(p)$ fermions to fill $p$ shells, we have the maps
$$ \eqalign{
\Lambda^{NA_N(p)} \TT^1_0(\lambda,w)
&\longrightarrow \TT^{NB_N(p)+NA_N(p)}_0(NA_N(p) \lambda, NA_N(p) w) \cr
&\longrightarrow \TT^0_0(A_N(p)(N\lambda + 1) + B_N(p), NA_N(p) w)
}	\eqno(4.47) $$
Similarly, if we start from a contravariant vector field,
$$ \eqalign{
\Lambda^{NA_N(p)} \TT^0_1(\lambda,w)
&\longrightarrow \TT^{NB_N(p)}_{NA_N(p)}(NA_N(p) \lambda, NA_N(p) w) \cr
&\longrightarrow \TT^0_0(A_N(p)(N\lambda - 1) + B_N(p), NA_N(p) w)
}	\eqno(4.48) $$

Another generalization deals with the recently\refto{11} discovered
$Vect(N)$ representations which we called {\it conformal fields},
because they transform naturally under the ``conformal'' subalgebra
$sl(N+1) \subset Vect(N)$.  The generators, in a form analogous to
(4.12), are
$$ L(f_\mu) =  u^{-1}(x) \Big( f_\mu(x) \d^\mu
+ (\d^B + k^B) f_\mu(x) T^\mu_B
+ c x_A \d^B \d^\mu f_\mu(x) T^A_B \Big) u(x),
	\eqno(4.49) $$
where $A, B = 0, 1, \ldots, N$ are ``conformal'' indices,
which take on $N+1$
different values. $T^A_B$ satisfies the algebra $gl(N+1)$,
$$ [T^A_B, T^C_D] = \delta^A_D T^B_C - \delta^B_C T^A_D,
	\eqno(4.50) $$
and
$$ \eqalign{
\d^A \equiv (\d^0, \d^\mu) &= (- x \cdot \d, \d^\mu), \cr
x_B \equiv (x_0, x_\nu) &= (1, x_\nu),	\cr
k^A \equiv (k^0, k^\mu) &= (1, 0).
}	\eqno(4.51) $$
Moreover, $c$ is a c-number parameter and $u(x)$ is the same arbitrary
function as in (4.12); on the torus, $u(x) = \exp(w \cdot x)$.

Denote the module with $p$ upper and $q$ lower conformal indices and
conformal weight $\lambda$ by $\CC^p_q(\lambda, c, w)$; when $w=0$ we
write
$\CC^p_q(\lambda, c)$. E.g., a conformal vector
transforms as
$$ \eqalign{
[-L(f_\mu), \psi^B(x)] &= f_\mu(x) (\d^\mu + w^\mu(x)) \psi^B(x)
+ \lambda \d^\mu f_\mu(x) \psi^B(x) \cr
&+ (\d^B + k^B) f_B(x) \psi^\mu(x)
+ c (\d^B \d^\mu f_\mu(x)) x_C \psi^C(x).
}	\eqno(4.52) $$

Scalar fields are a special case of both tensor fields and conformal
fields ($T^A_B = \lambda \delta^A_B$). It can be shown that there is a
map\refto{12}
$$ \eqalign{
\CC^0_0(\lambda, c, w) &\longrightarrow \CC^1_0(\lambda, c, w) \cr
(d\psi)^A &= (\d^A + w^A + \lc k^A) \psi
}	\eqno(4.53) $$
However, the ground state molecules constructed with help of this map
are essentially the same as above. To see this, consider the molecule
with one full shell.
$$ \psi \d^0 \psi \d^1 \psi \ldots \d^N \psi
= - x_\nu \psi \d^\nu \psi \d^1 \psi \ldots \d^N \psi
\equiv 0,
	\eqno(4.54) $$
because $\d^\nu \psi$ appears twice for every value of $\nu$. If we
start with non-scalar conformal fields, new molecules can be found.
Invariants are then obtained by contracting with the
$(N+1)$-dimensional permutation symbol, considered as the totally
skew, constant conformal field
$$ \epsilon_{A_0 A_1 \ldots A_N}(x) \in \CC^0_{N+1}(1, c),
	\eqno(4.55) $$
or, dually,
$$ \epsilon^{A_0 A_1 \ldots A_N}(x) \in \CC_0^{N+1}(-1, c).
	\eqno(4.56) $$
That this definition is consistent is established analogously to
(4.30). It should be noted that the value of $c$ is arbitrary, because
it will only enter the transformation law multiplied by the factor
$$ x_B \d^B \d^\mu f_\mu(x) \epsilon_{A_0 A_1 \ldots A_N}(x) \equiv 0.
	\eqno(4.57) $$
For example, there are maps
$$ \eqalign{
\Lambda^{(N+1)A_N(p)} &\CC^1_0(\lambda,c,w) \cr
\longrightarrow
&\CC^{(N+1)B_N(p)+(N+1)A_N(p)}_0((N+1)A_N(p) \lambda,
c, (N+1)A_N(p) w) \cr
\longrightarrow
&\CC^0_0(A_N(p)((N+1)\lambda + 1) + {{(N+1)B_N(p)} \over N},
c, (N+1)A_N(p) w) .
}	\eqno(4.58) $$
Using the fact that $\CC^0_0(\lambda, c, w) = \TT^0_0(\lambda, w)$,
it is now
straightforward to check when these maps give invariants.

\endchapter

\header{5. Fock modules}

The first part of the Feigin-Fuks procedure, the construction of
invariants, was generalized to higher dimensions in the previous
section. The second step consists of applying these invariants to the
Fock vacuum in order to obtain singular vectors. Unfortunately, the
construction of Fock modules is technically and conceptually difficult
for $N > 1$, because infinities arise that can not be removed by
normal ordering. The problem is essentially that subspaces of fixed
energy are infinite-dimensional. In this section we discuss this issue
and indicate a way out of the mathematical difficulty; however, it is
not the physically relevant solution.

We first explain why it is not possible to find representations on the
unconstrained Fock space.  Let the fermion $\psi(x)$ and its canonical
conjugate $\psic(x)$ satisfy CAR,
$$ \eqalign{
\{\psic(x), \psi(y)\} &= \one \delta(x-y) \cr
\{\psic(m), \psi(n)\} &= \one \delta(m+n)}
\qquad
\eqalign{
\{\psic(x), \psic(y)\} &= \{\psi(x), \psi(y)\} = 0, \cr
\{\psic(m), \psic(n)\} &= \{\psi(m), \psi(n)\} = 0. }
	\eqno(5.1) $$
Any tensor, conformal or internal indices are suppressed and $\one$ is
the unit matrix.

As in Sec. 2, we consider $\psi(x)$ as coordinates and $\psic(x)$ as
derivatives on a vector space $V_\psi$, and the envelopping algebra of
(5.1) is the fermionic Weyl algebra on this space.  $Vect(N)$ acts (by
commutation) as the first-order differential operators
$$ \eqalign{
L^\mu(m) &= \int d^Nx \; \emx \; \psic(x)
\Big( (\d^\mu + w^\mu) \psi(x) + m^\sigma [T^\mu_\sigma, \psi(x)] \Big) \cr
&= \sum_s \Big( (-s^\mu + w^\mu) \psic(m-s) \psi(s)
+ m^\sigma \psic(m-s) [T^\mu_\sigma, \psi(s)] \Big),
}	\eqno(5.2) $$
where the sum runs over $s \in \Lambda_\psi$, where $\Lambda_\psi$ is
obtained from the lattice $\Lambda$ by a constant translation
(analogous to $1/2$ in one dimension).  If $\psi \in
\TT^p_q(\lambda,w)$, we read off from (5.2) that its conjugate
transforms as $\psic \in \TT^q_p(1-\lambda,-w)$.  The problem of
representing $Vect(N)$ has thus essentially been reduced to the
appearently simpler problem of representing the Weyl algebra.  This
intuition is false, and even more so than in one dimension, due to the
infinite dimensionality of $Vect(N)$.

To construct a Fock module, we must choose a polarization which
separates the elements into two sets: raising and lowering operators.
One possibility is to divide the lattice into two parts:
$\Lambda_\psi = \Lambda^{(+)}_\psi \oplus \Lambda^{(-)}_\psi$.
We write $m>0$ ($m<0$) if $m \in \Lambda^{(+)}_\psi$
($m \in \Lambda^{(-)}_\psi$).
The decomposition defines an order of the lattice points provided that
$m,n>0$ implies that $m+n > 0$ and $-m < 0$. For example, introduce a
constant vector $t_\mu$, say $t_\mu = (0, \ldots, 0, 1)$, and proclaim
that $m > 0$ ($m < 0$) if $t_\mu m^\mu > 0$ ($t_\mu m^\mu < 0$).  This
procedure can be expressed physically as follows. Introduce a
{\it Hamiltonian}
$$ H = t_\mu L^\mu(0) = L^N(0),
	\eqno(5.3) $$
that generates rigid translations in the ``time'' direction $N$.
Every Fourier component of the field has a definite energy, because
$$ [H, \psi(n)] = t_\mu n^\mu \psi(n) .
	\eqno(5.4) $$
The vector space associated to the field thus has a decomposition
into energy eigenspaces,
$$ V_\psi = \bigoplus_{j = -\infty}^\infty V_\psi^{(j)}
= V_\psi^{(+)} \oplus V_\psi^{(-)}.
	\eqno(5.5) $$
A Fock module is now defined by a vacuum state $\vac$, defined by
the relations
$$ \eqalign{
\psi(-n) \vac &= \psic(-n) \vac = 0,
\qquad \forall n > 0 \cr
}	\eqno(5.6) $$
This is a representation of the Weyl algebra, but
the inherited $Vect(N)$ representation has an infinite central extension.
For a scalar field, the relevant calculation is
$$ [L^\mu(m), L^\nu(-m)] \vac = \sum_{0 < s < m}
(-s^\mu + \lambda m^\mu + w^\mu) (s^\nu + (\lambda-1) m^\nu - w^\nu) \vac
	\eqno(5.7) $$
and this sum is infinite except in one dimension, because there are
infinitely many points perpendicular to the ``time'' direction. An
infinite extension is definitely not acceptable, so some additional
idea has to be invoked.

The Hamiltonian (5.3) was introduced on purely mathematical grounds to
generate a $\Ints$-gradation by energy.  A more natural Hamiltonian is
the scaling operator $H = x_\mu \d^\mu$, which corresponds to a
polarization by the radial component.  Express an arbitrary vector
field in spherical coordinates.
$$ L(f_\mu) = \sum_{n = -\infty}^\infty \sum_{l=0}^\infty \sum_{m=-l}^l
f_{\mu,nlm} L^\mu_{nlm} r^n Y_{lm}(\theta, \varphi).
	\eqno(5.8) $$
We have specialized to three dimensions, but (5.8) is readily
generalized to arbitrary dimension by considering hyper-spherical
harmonics. The energy is now given by the radial power $n$, and the
algebra $Vect(N) \equiv \LL$ has a decomposition in homogeneous
components
$$ \LL = \bigoplus_{n = -\infty}^\infty \LL^{(n)}
= \LL^{(+)} + \LL^{(-)},
	\eqno(5.9) $$
where
$\{L^\mu_{nlm} : l \in [0,\infty[, m \in [-l, l], \mu \in [1, N] \}$
is a basis for $\LL^{(n)}$. The vacuum obeys $\LL^{(-)} \vac = 0$, i.e.
$$ L^\mu_{nlm} \vac = 0,
	\eqno(5.10) $$
for every $n < 0$, and for every $\mu$, $l$ and $m$.

$\LL^{(+)}$ is the algebra of vector fields which are non-singular in
some neighborhood of the origin; because of general covariance, any
sensible object must be a representation of it. Schematically,
$$ [\LL^{(+)}, \LL^{(+)}] \subset \LL^{(+)} .
	\eqno(5.11) $$
Similarly, the elements of $\LL^{(-)}$ are non-singular close to infinity.
$$ [\LL^{(-)}, \LL^{(-)}] \subset \LL^{(-)} .
	\eqno(5.12) $$
However, $\LL^{(+)}$ and $\LL^{(-)}$ are nowhere simultaneously
non-singular.  Commutators involving both subalgebras can not be
trusted to have their classical form. Rather, we expect that
$$ [\LL^{(+)}, \LL^{(-)}] \subset \LL^{(+)} \oplus \LL^{(-)} \oplus
\hbox{anomaly}.
	\eqno(5.13) $$
In one dimension, the anomaly (Schwinger term) is simply the central
extension of the Virasoro algebra.  The presence of this anomaly does
not really violate general covariance, because a singular vector field
is not truly an infinitesimal coordinate transformation.

We na\"\i vely expect to obtain a module of the type above by
expanding the
fermion field in spherical coordinates,
$$ \psi(x) = \sum_{n = -\infty}^\infty \sum_{l=0}^\infty \sum_{m=-l}^l
\psi_{nlm} r^n Y_{lm}(\theta, \varphi),
	\eqno(5.14) $$
and similar for $\psic(x)$, and inserting these expressions into
(5.2).  However, it is easy to see that we pick up an infinite anomaly
for precisely the same reason as with the Fourier polarization (5.3):
every subspace of fixed energy is infinite-dimensional,
$$ \dim V_\psi^{(n)} =  \sum_{l=0}^\infty \sum_{m=-l}^l 1 = \infty.
	\eqno(5.15) $$

The same kind of infinities arise also in other bases, and in other
algebras that act on $N$-dimensional space, e.g. $Map(N, \oj)$.  The
problem is thus generic. If we regard the presence of an infinite
anomaly as a fundamental inconsistency, as we do, the conclusion must
be that these quantized fields do not make sense except in one
dimension. In the remainder of this paper we discuss various means to
deal with this problem.

Physically, we might expect to remove the infinities by
renormalization. Normal ordering, which has a natural role in
representation theory, is equivalent to mass renormalization, but in
physics there is also wave-function renormalization.  However, it is
not clear to us how renormalization ideas can be implemented within
our formalism. Moreover, we are dealing with the diffeomorphism group,
which is intimately linked to unrenormalizable gravity.

Related ideas been investigated by Mickelsson and collaborators in the
context of current algebras, extending the Pressley-Segal approach to
loop groups to higher dimensions.\refto{17-20} Eq. (5.2) defines an
embedding of $Vect(N)$ in $GL(V_\psi)$, the group of linear
transformations of $V_\psi$. One can define various restrictions of
$GL(V_\psi)$, and attempt to represent the algebra on these
restrictions.

In Ref. 11 we proposed to introduce fundamental bosons, because they
yield infinities of opposite sign. By matching fermionic and bosonic
degrees of freedom, the infinities would cancel. This is not a
satisfactory solution, because the bosonic and fermionic sectors
decouple, and hence such a module would decompose into a direct sum of
two unacceptable modules.  If one could introduce interaction terms in
the $Vect(N)$ generators, e.g. a term which destroys a boson and
creates two fermions, the different sectors would not decouple, and
the matching argument might deserve further attention. However, we
have found no way to achieve this.

As an alternative to the approaches above, we propose to circumvent
the problem in a mathematically very simple manner. Eq. (5.2) defines
an action of $Vect(N)$ on the vector space $V_\psi$, but this vector
space is too large. By imposing some constraints on the fields, we can
restrict the action to some submanifold (the {\it constraint
manifold}). The constraints must evidently transform covariantly under
arbitrary diffeomorphisms, to make the submanifold stable under the
action of $Vect(N)$. If such a constraint is found, we automatically
get an action on the function and Weyl algebras on the constraint
manifold, and these algebras might be small enough to admit a
$\Ints$-gradations by finite-dimensional subspaces. If so, we can
reduce the representations further by introducing a vacuum that
annihilates all negative components.  The basic idea is hence that a
free field has too many degrees of freedom, but the constraints cut
down the size of the field enough to make it managable.

It should be noted that the idea of imposing constraints is
mathematically quite inevitable, because our goal is to construct
irreducible representations. Consider some set of fields as a
$Vect(N)$ module.  If it is possible to write down a non-trivial
covariant equation, the module decomposes into the solution space and
its complement, both of which are submodules.  Hence, unless one of
these submodules is trivial (and the other is the original module), we
have managed to reduce our module. For example, a metric can be
considered as the $Vect(N)$ module $\TT^2_0(0)$, which can be
decomposed into one submodule with and one without curvature. From
this point of view, every conceivable equation must be satisfied.

To illustrate that constraints may lead to finite anomalies, we
consider a fermion satisfying the massless Dirac equation. There is an
obstacle here because $Vect(N)$ has no natural spinor representation.
To deal with this we introduce the {\it frame algebra} $Map(N, so(N))$
as in (4.39).
$$
[J^{ij}(m), J^{kl}(n)]
= \eta^{ik} J^{jl}(m+n) - \eta^{il} J^{jk}(m+n)
- \eta^{il} J^{jk}(m+n) + \eta^{jl} J^{ik}(m+n).
	\eqno(5.16) $$
There is no distinction between upper and lower frame indices, due to
the constant metric $\eta_{ij}$, and we will therefore mix them
freely.  A spinor is defined to transform as the scalar $\TT^0_0(0)$
under $Vect(N)$ and as a spinor under the frame algebra.  The total
algebra has thus been enlarged to $Vect(N) \IX Map(N, so(N))$, and we
want to construct representations of this larger algebra.

The massless Dirac equation reads
$$ \gamma_i e^i_\mu(x) (\d^\mu + \omega^\mu(x)) \psi(x) = 0
	\eqno(5.17) $$
Here $\gamma_i$ are the constant gamma matrices,
$\sigma_{ij} = [\gamma_i, \gamma_j]/4i$,
and $e^i_\mu(x)$ is a vielbein field,
which is a frame vector transforming as $\TT^0_1(0)$ under $Vect(N)$.
Further, the vielbein has an inverse
$$ e^i_\mu(x) e_j^\mu(x) = \delta^i_j,
\qquad
e^i_\nu(x) e_i^\mu(x) = \delta^\mu_\nu,
	\eqno(5.18) $$
and $\omega^\mu(x) = \omega^{i\mu}_j(x) \sigma^{ij}$ is the
spin connection,
which is defined in terms of the vielbein by
$$ \d^{[\mu} e^{\nu]}_i + \omega_i^{j[\mu} e^{\nu]}_j = 0.
	\eqno(5.19) $$

To solve (5.17) for $\psi(x)$ we adopt a special coordinate system in which
$$ e^i_\mu(x) = \delta^i_\mu,
\qquad \omega^{i\mu}_j(x) = 0
	\eqno(5.20) $$
It must be emphasized that this choice is not covariant and thus not
really acceptable, but it serves to illustrate our point. Using
(5.20), we have $\gamma_i \d^i \psi(x) = 0$, which as usual implies
that the components satisfy the Laplace equation. Specializing to
spherical coordinates in three dimensions, the general solution reads
$$ \psi(x) = \sum_{l=0}^\infty \sum_{m=-l}^l
\Big( \psi^{(+)}_{lm} r^l +  \psi^{(-)}_{lm} r^{-l-1})
Y_{lm}(\theta, \varphi).
	\eqno(5.21) $$
The gradation is given by
$$ \deg \psi^{(+)}_{lm} = l,
\qquad \deg \psi^{(-)}_{lm} = -l-1.
	\eqno(5.22) $$
and the dimension of the homogeneous subspace
$$ \dim V_\psi^{(l)} = \dim V_\psi^{(-l-1)}
=  \sum_{m=-l}^l 1 = 2 l + 1
	\eqno(5.23) $$
is finite. Similarly, $\dim V_\psi^{(l)} \propto |l|^{N-2}$
for large $|l|$ in $N$ dimensions.
The coefficients $\psi^{(\pm)}_{lm}$ are coordinates
on the constraint manifold and
the conjugate variables $\psi^{(\pm)\dagger}_{lm}$,
$$ \{\psi^{(\pm)\dagger}_{lm}, \psi^{(\mp)}_{l\pr m\pr} \}
= \delta_{l l\pr} \delta_{m m\pr},
	\eqno(5.24) $$
are the corresponding tangential derivatives. Since the constraint
manifold is stable, we can express the $Vect(N)$ generators in terms
of these conjugate variables, essentially as in (5.2).  We can now
define the vacuum by
$$ \psi^{(-)}_{lm} \vac = \psi^{(-)\dagger}_{lm} \vac = 0.
	\eqno(5.25) $$

A serious flaw in the above argument is that the equation (5.20) is
not covariant, wherefore the fermions can not invariantly be expanded
in solutions to the flat Laplace equation.  However, for continuity
reasons we expect the expansion (5.21) to be almost correct in
coordinate systems close to (5.20). A smooth deformation of (5.21) is
$$ \psi(x) = \sum_{l=0}^\infty \sum_{m=-l}^l
\Big( \psi^{(+)}_{lm} F_{lm}(r, \theta, \varphi; e) +
\psi^{(-)}_{lm} F_{-l-1,m}(r, \theta, \varphi; e) \Big) ,
	\eqno(5.26) $$
where the Laplace eigenfunctions $ F_{lm}(r, \theta, \varphi; e)$
depend on the vielbein, as does the vacuum $\state{0; e^i_\mu(x)}$.
In the case of a diagonal vielbein,
$$ F_{lm}(r, \theta, \varphi; \delta^i_\mu) = r^l Y_{lm}(\theta, \varphi)
	\eqno(5.27) $$
Both the eigenfunctions and the vacuum must transform non-trivially
under $Vect(N)$ because of this dependence, but the number of
eigenfunctions, and thus the dimensions of the spaces $V_\psi^{(l)}$,
can not change continuously. This makes it plausible that these
dimensions remain finite for all vielbeins, at least in some suitable
class, and consequently the anomaly does not diverge. We have thus
defined, modulo these technical assumptions about continuity, a
$Vect(N)$ module which is partially of lowest-weight type.

An analogous module has been studied by Mickelsson and Rajeev in the
gauge case, where the role of the vielbein is played by a gauge
potential.\refto{17-18} It appears that they have encountered serious
problems (non-unitarizability) with this approach.  It is quite clear
that this module is not what we want: the vielbein field is still
classical, i.e. not of lowest-weight type. To remedy this, consider
the Weyl algebra built out of both vielbeins and fermions and their
conjugates.  To (5.1) we add the canonical commutation relations,
$$ [e_j^\mu(x), e^{i\dagger}_\nu(y)]
= \delta^\mu_\nu \delta^i_j \delta(x-y),
\qquad
[e^{i\dagger}_\mu(x), e^{j\dagger}_\nu(y)]
= [e_i^\mu(x), e_j^\nu(y)] = 0.
	\eqno(5.28) $$
The vielbein contribution to $Vect(N)$,
$$ L_e^\mu(m) = \int d^Nx \; \emx \; e^{j\dagger}_\nu(x)
\Big( \d^\mu e_i^\nu(x) + m^\nu e_i^\mu(x) \Big),
	\eqno(5.29) $$
is added to (5.2) to get the total generator.

We now introduce the Dirac equation (5.17) and consider the Weyl
algebra on the constraint manifold. It is a $Vect(N)$ module, with
action given in principle by (5.2) and (5.29).  Unfortunately, the
description of this module is severely complicated by the fact that
the constraint is non-linear in the vielbein. Previously, the
constraint was linear in $\psi(x)$, although it was parametrized by
the classical field $e^i_\mu(x)$, and the submanifold was a family of
vector spaces. We could therefore write down the general solution as a
sum of eigenfunctions.  This is no longer the case.

Another complication is that the quantized vielbein will give rise to
the same kind of infinite anomalies as the unconstrained fermion did
above. To see this, note that the constraint is preserved by the
fermionic number operator,
$$ [N_\psi, \hbox{constraint}] = \hbox{constraint}.
	\eqno(5.30) $$
We can thus limit our attention to modules with a fixed number of
fermions. The zero-fermion sector is completely unconstrained because
the Dirac equation is an identity there. The analysis that led to
(5.7) remains unchanged except that the fermion is replaced by a
boson, and an infinite anomaly arises.

In view of the previous development it is now clear how to proceed:
introduce another constraint that cuts down the number of degrees of
freedom, to obtain an appropriate $\Ints$-gradation. The constraint
must transform covariantly and it can only depend on the vielbein
since we are in the zero-fermion sector. Moreover, it should be as
restrictive as possible without making the vielbein trivial. The
simplest conceivable equations thus involve the curvature, or some of
its derivates, such as the Ricci or Einstein tensors.

The strategy for constructing more complicated modules is now clear;
find a suitable set of constraints to impose on some Weyl algebra. A
constraint should preferably be as simple as possible, for maximal
reduction of the degrees of freedom. On the other hand, it must not be
so restrictive as to trivialize any field. We presumably strike a
sensible balance by considering first- and second-order differential
equations.

For a scalar field $\psi(x)$ with non-zero conformal weight $\lambda$,
the Dirac equation has to be modified, because the gradient of
$\psi(x)$ is no longer a tensor field. The modified Dirac equation
reads
$$ \gamma_k e^k_\mu(x)
(\d^\mu + \lambda \Gamma^{\sigma\mu}_\sigma(x)
+ \omega^{i\mu}_j(x) \sigma_{ij}) \psi(x)
= i M \psi(x),
	\eqno(5.31) $$
where
$$ \Gamma^{\sigma\mu}_\tau(x) =
e^\sigma_i(x) (\d^\mu \delta^i_j + \omega^{i\mu}_j(x) ) e^j_\tau(x)
	\eqno(5.32) $$
is the Christoffel symbol and we have introduced a non-zero constant
mass $M$.  Eq. (5.31) is manifestly covariant both under $Vect(N)$ and
$Map(N, so(N))$ and is hence a meaningful equation.

The number of fundamental fields can be increased further by the
introduction of a Yang-Mills field. Every field must then transform
consistently under some additional gauge algebra (4.39), in addition
to the frame algebra $so(N)$ (or $so(N+1)$ for conformal fields). The
equations must be covariantized with respect to this extra gauge
algebra, which amounts to the introduction of the gauge potential in
the Dirac equation and in the energy-momentum tensor.  If the
Yang-Mills field were free it would generate an infinite Schwinger
term by the same mechanism as before. This is avoided by demanding
that the field strength satisfy the Yang-Mills equation with some
current constructed from the fermions.

\endchapter

\header{6. Conformal fields }

The maximally constrained modules of the previous section were
mathematically appealing: they were of lowest-weight type, presumably
had a finite anomaly, and could not be further reduced by imposing
equations. However, they are not the modules of interest in physics.
Classical physics fields obey certain equations of motion in
space-time, but the fields and momenta on a space-like surface are
freely specifyable. On the other hand, a mathematically sensible
action of $Vect(N)$ on the quantized phase space is only possible if
the fields obey some equations already at fixed time. The fields must
thus obey two fundamentally different types of equations.

Or must they? With the recent discovery of conformal fields, a new
possibility opens up, which roughly can be described as projecting the
time derivative onto space derivatives.  We can then take this
``time'' derivative, substitute it into conformal analogs of the
classical equations of motion, and obtain kinematical constraints on
the quantum fields at fixed time. This idea of trading dynamics for
kinematics may admittedly sound absurd, but it does make mathematical
sense, as we now proceed to show.

In the plane-wave basis, the conformal fields (4.49) are defined by
$$ L^\mu(m) = \emx \Big( \d^\mu + (m^A + k^A) T^\mu_A
+ c m^\mu m^A T^B_A x_B \Big),
	\eqno(6.1) $$
where the $(N+1)$-dimensional coordinate is
$$ x_B = (t, x_\nu),
	\eqno(6.2) $$
with $t$ being an arbitrary constant. Further,
$T^A_B$ generate $gl(N+1)$, i.e.
$$ [T^A_B, T^C_D] = \delta^A_D T^C_B - \delta^C_B T^A_D,
	\eqno(6.3) $$
and
$$
k^A = (t^{-1}, 0), \qquad
\d^A = (- t^{-1} x \cdot \d, \d^\mu), \qquad
m^A = (- t^{-1} x \cdot m, m^\mu),
	\eqno(6.4) $$
The introduction of the parameter $t$ makes (6.4) slightly more
general than (4.51) and the expression in Ref. 11. However, it is
readily verified that (6.2) and (6.4) still obey the algebraic
relations
$$
m^A x_A = x_A \d^A = 0, \qquad
k^A x_A = 1,
	\eqno(6.5) $$
and
$$ \eqalign{
[\d^A, \d^B] &= k^A \d^B - k^B \d^A \cr
[\d^A, x_B] &= \delta^A_B - k^A x_B
} \qquad \eqalign{
[\d^A, m^B] &= - k^B m^A \cr
[\d^A, \emx] &= m^A \emx,
}	\eqno(6.6) $$
and all commutators between $m^A$, $x_B$ and $k^C$ vanish.
The proof that (6.1) satisfies $Vect(N)$ is repeated in the appendix.

{}From the expression for $x_B$ we see that conformal fields are
naturally equipped with a length scale $t$, so in that sense their
name might be poorly chosen. We will think of this parameter as
``time'', which is fixed on the space-like surface where $Vect(N)$
acts. The units are such that the velocity of light is one.  Of
course, the algebraic structure is independent of this physical
interpretation, but it is useful at least as a naming convention. Any
conformal field can be divided into time and space components. E.g.,
if we split a conformal vector as
$$ \psi^A(x) = (\phi(x), \psi^\nu(x)),
	\eqno(6.7) $$
the transformation law
$$ [L^\mu(m), \psi^A(x)]
= -\emx \Big( (\d^\mu + \lambda m^\mu) \psi^A(x) +
(m^A + k^A) \psi^\mu(x) + c m^\mu m^A x_B \psi^B(x) \Big)
	\eqno(6.8) $$
takes the form
$$ \eqalign{
[L^\mu(m), \phi(x)]
&= -\emx \Big( (\d^\mu + \lambda m^\mu) \phi(x) +
t^{-1} (-m \cdot x + 1) \psi^\mu(x) \cr
&\qquad - c t^{-1} m^\mu m \cdot x
(t\phi(x) + x_\nu \psi^\nu(x)) \Big) \cr
[L^\mu(m), \psi^\nu(x)]
&= -\emx \Big( (\d^\mu + \lambda m^\mu) \psi^\nu(x) +
m^\nu \psi^\mu(x) + c m^\mu m^\nu (t \phi(x) + x_\nu \psi^\nu(x)) \Big).
}	\eqno(6.9) $$
Similarly,
$$ [L^\mu(m), \psi_B(x)] = -\emx \Big( (\d^\mu + \lambda m^\mu) \psi_B(x)
- \delta^\mu_B (m^A + k^A) \psi_A(x) - c m^\mu m^A x_B \psi_A(x) \Big)
	\eqno(6.10) $$
becomes
$$ \eqalign{
[L^\mu(m), \phi(x)] &= -\emx \Big( (\d^\mu + \lambda m^\mu) \phi(x)
- c t m^\mu (- t^{-1} m \cdot x \phi(x) + m^\nu \psi_\nu(x)) \Big) \cr
[L^\mu(m), \psi_\nu(x)]
&= -\emx \Big( (\d^\mu + \lambda m^\mu) \psi_\nu(x)
- \delta^\mu_\nu \big( t^{-1} (- m \cdot x + 1) \phi(x)
+ m^\sigma \psi_\sigma(x) \big) \cr
&\qquad - c m^\mu (- t^{-1} m \cdot x \phi(x) + m^\nu \psi_\nu(x)) \Big),
}	\eqno(6.11) $$
when divided into time and space components:
$$ \psi_B(x) = (\phi(x), \psi_\nu(x)).
	\eqno(6.12) $$

The time parameter can be eliminated by replacing $\phi(x)$ by $t
\phi(x)$ in both (6.7) and (6.12). However, it is useful for showing
that a tensor field in a sense is a special case of a conformal field.
Set $c = 0$ and let $t \to \infty$ in (6.1). In this limit, $k^A \to
0$ and $m^A \to m^\mu$, so
$$ L^\mu(m) \to \emx(\d^\mu + m^\sigma T^\mu_\sigma),
	\eqno(6.13) $$
which defines a tensor field. In this limit, a conformal
vector decomposes into
a direct sum of its time and space components,
$$ \CC^1_0(\lambda, 0) \to \TT^0_0(\lambda) \oplus \TT^1_0(\lambda).
	\eqno(6.14) $$
A generic conformal field with $c \ne 0$ does not correspond
to a tensor field.

We have recently studied intertwining operators that connect different
conformal fields.\refto{12} First, conformal fields with the same
value of $c$ can be multiplied, because Leibniz' rule holds both for
the derivative $\d^A$ and for the $gl(N+1)$ generators $T^A_B$.
Second, there are two types of first-order differential operators.
The first involves the totally skew {\it positive conformal forms}
$\Omega^p(\lambda,c) \subset \CC^p_0(\lambda,c)$.
$$ \eqalign{
d_p(\lambda,c): \quad
\Omega^p(\lambda, c) &\longrightarrow \Omega^{p+1}(\lambda, c) \cr
(\phi_p)^{A_1 \ldots A_p}
&\mapsto (d_p(\lambda,c)\phi_p)^{A_1 \ldots A_{p+1}}
\equiv {1 \over {(p+1)!}} (\d^{[A_1} + \gamma_p(\lambda,c) k^{[A_1})
(\phi_p)^{A_2 \ldots A_{p+1}]}
}	\eqno(6.15) $$
where $\gamma_p(\lambda,c) = \lambda/c - p$.
The second map involves the totally skew {\it negative conformal forms}
$\Omega_p(\lambda,c) \subset \CC^0_p(\lambda,c)$.
$$ \eqalign{
\dn_p(\lambda,c): \quad
\Omega_p(\lambda, c) &\longrightarrow \Omega_{p-1}(\lambda, c) \cr
(\phi_{-p})_{A_1 \ldots A_p}
&\mapsto (\dn_p(\lambda,c)\phi_{-p})_{A_1 \ldots A_{p-1}}
\equiv (\d^B + \gamman_p(\lambda,c) k^B) (\phi_{-p})_{A_1 \ldots A_{p-1}B}
}	\eqno(6.16) $$
where $\gamman_p(\lambda,c) = (\lambda-1)/c + p - N - 1$.

These maps satisfy $d_{p+1} d_p = 0$ and $\dn_{p-1} \dn_p = 0$, with
the exception $d_0 \dn_1 \ne 0$.  This is the natural generalization
of the exterior derivative to conformal fields, but it should be noted
that these {\it conformal exterior derivatives} are much more abundant
than the usual ones; they can be defined for any value of the
conformal weight $\lambda$.

With these tools we can now immediately write down conformal
generalizations of most equations in physics. The recipe is simply to
substitute
$$ {{\d} \over {\d t}} \to \d^0 + \lc k^0
= - x \cdot \d + {\lambda \over {ct}}
	\eqno(6.17) $$
in derivatives acting on scalar fields.  When acting on non-scalar
conformal fields, this recipe has to be modified in accordinance with
(6.15-16). This is what we mean by ``projecting the time derivative
onto space''. We stress that this procedure does make sense, because
the conformal exterior derivatives are invariantly defined.

The simplest conceivable conformal equation involves a single scalar
field $\phi(x)$.
$$ (d\phi)^A(x) \equiv (\d^A + \lc k^A) \phi(x) = 0.
	\eqno(6.18) $$
Multiplying with $x_A$ we find that $\lambda \phi(x) = 0$, i.e. either
$\lambda = 0$ or $\phi(x) \equiv 0$. In the former case, the space
components of (6.18) read $\d^\nu \phi(x) = 0$. The only non-zero
solution is thus that $\phi(x)$ is constant, which is too restrictive.

To write down covariant equations with non-trivial solutions, we must
introduce more conformal fields.  A {\it conformal vielbein}
$e^A_i(x)$, which is a $\CC^1_0(0,c)$ field transforming as a vector
under the frame algebra $Map(N, so(N+1))$.  This is of course the same
algebra as (5.16), apart from the dimension of the frame vectors.  The
vielbein is defined by the non-trivial property of having a two-sided
inverse $e^i_A(x)$ everywhere.
$$ e^A_i(x) e^i_B(x) = \delta^A_B, \qquad
e^A_j(x) e^i_A(x) = \delta^i_j.
	\eqno(6.19) $$
It is this property that selects the gauge group $so(N+1)$ ($sl(N+1)$
would also be possible, but it has no spinor representations).
Clearly, the inverse vielbein transforms as $\CC^0_1(0,c)$ and as an
$so(N+1)$ vector.

A {\it conformal Laplace equation} reads
$$  e^i_A(x) \Big(\d^A \eta_{ij} + \lc k^A \eta_{ij}
+ \omega^A_{ij}(x) \Big)
 e^j_B(x) \Big(\d^B + \lc k^B \Big) \phi(x) = 0
	\eqno(6.20) $$
where the {\it conformal spin connection}
$\omega^A_{ij} \in \CC^A_0(0, c)$
transforms as a connection under the frame algebra, i.e.
$$ [J_{ij}, \omega^B_{kl}(x)]
= \emx \Big( \eta_{ik} \omega^B_{jl}(x) -  \eta_{il} \omega^B_{jk}(x)
 - \eta_{jk} \omega^B_{il}(x) +  \eta_{jl} \omega^B_{ik}(x)
+ m^B \eta_{ij} \Big).
	\eqno(6.21) $$
The spin connection can as usual be defined in terms of the vielbein
by the condition of vanishing torsion,
$$ \d^{[A} e^{B]}_i(x) + \Big(\lc - 1 \Big) k^{[A} e^{B]}_i(x)
+ \omega_i^{j[A}(x) e^{B]}_j(x) = 0.
	\eqno(6.22) $$
That (6.20) and (6.22) are covariant both under $Vect(N)$ and
$Map(N, so(N+1))$ follows immediately from the existence of the conformal
exterior derivatives. The conformal Laplace equation is actually
somewhat simpler than its standard counterpart, which would require
the use of Christoffel symbols for non-zero $\lambda$.

Similarly, we can write down a {\it conformal Dirac equation} if $\psi(x)$
is a frame spinor.
$$ \gamma_i e^i_A(x)  \Big(\d^A + \lc k^A + \omega^A(x) \Big) \psi(x)
= i M \psi(x),
	\eqno(6.23) $$
where $\gamma_i$ and $\sigma_{ij}$ are the $so(N+1)$ gamma and
spin matrices and
$$ \omega^A(x) = \omega^{iA}_j(x) \sigma_{ij}.
	\eqno(6.24) $$
The number of fermionic degrees of freedom can be halved when $M=0$ by
considering chiral spinors. The conformal Dirac equation (6.23) is
nevertheless reminiscent of a massive equation, because in flat space
it has the structure
$$ \Big(\gamma_i \d^i + \lc \gamma^0 \Big) \psi(x) = 0,
	\eqno(6.25) $$
and the role of mass is played by the parameter $\lambda/c$.

Skew conformal fields may be dualized in the same fashion as tensor fields.
A volume element can be defined as
$$ v = e^{A_0}_{i_0} e^{A_1}_{i_1} \ldots e^{A_N}_{i_N}
\epsilon_{A_0 A_1 \ldots A_N}  \epsilon^{i_0 i_1 \ldots i_N}
\in \CC^0_0(1,0),
	\eqno(6.26) $$
where the transformation law of the permutation symbol (4.55) was used.
A dual conformal field is defined by contraction with one of
$$ \eqalign{
E_{A_0 A_1 \ldots A_N}
&= v^{-1} \epsilon_{A_0 A_1 \ldots A_N} \in \CC^0_{N+1}(0, c), \cr
E^{A_0 A_1 \ldots A_N}
&= v \epsilon^{A_0 A_1 \ldots A_N} \in \CC_0^{N+1}(0, c), \cr
}	\eqno(6.27) $$
This notion of duality allows us to write down conformal analogs of
the Einstein equation.  The conformal spin connection has the
curvature
$$ R_{ij}{}^{AB} = \bigg(\d^{[A} + \Big( \lc-1 \Big) k^{[A} \bigg)
\omega^{B]}_{ij} + \omega^{[A}_{ik} \omega^{B]}_{kj},
	\eqno(6.28) $$
i.e.
$$ R_{CD}{}^{AB} = e^i_C e^j_D R_{ij}{}^{AB},
	\eqno(6.29) $$
and the Einstein tensor is the contraction of the double dual of (6.29).

The preceding paragraphs illustrate that it is possible to construct
conformal analogs of most equations in differential geometry (see also
Ref. 12). We will not pursue this topic further, but hope to return to
it in a forthcoming publication.  In the remainder of this section we
discuss the choice of Hamiltonian.  It is tempting to extend the
definition (6.1) by equipping $L^\mu(m)$ with a time component,
$$ L^A(m) = \emx \Big( \d^A + (m^B + k^B) T^A_B + c m^A m^B T^C_B x_C \Big).
	\eqno(6.30) $$
A slightly more general expression is also possible.\refto{12} Since
the space components $L^\mu(m)$ generate diffeomorphisms in space, it
is natural to think of the new time components as the generators of
diffeomorphisms in the time direction, and the Hamiltonian can be
identified as the generator of rigid time translations, $ H = L^0(0)$.
{}From
$$ H = \d^0 + k^B T^0_B = - x_\mu \d^\mu - T^\mu_\mu
= - {{\d L^\mu(m)} \over {\d m^\mu}} \Bigg|_{m = 0} ,
	\eqno(6.31) $$
we see that our Hamiltonian in fact is the dilatation operator. This
presumably indicates that the natural $\Ints$-gradation is by the
radial exponent, at least for modules built from conformal fields.

Equivalently, the Hamiltonian acts by commutation
on the fermionic Weyl algebra as the operator
$$ \eqalign{
H_\psi &= \int d^Nx \; \psic(x)
\Big( x_\mu \d^\mu \psi(x) +  [T^\mu_\mu, \psi(x)] \Big) \cr
&= \int d^Nx \; \psic(x) (x_\mu \d^\mu - \lambda) \psi(x) ,
}	\eqno(6.32) $$
where the last expression holds for $\psi(x) \in \CC^0_0(\lambda,c)$.
Eq. (6.32) is bilinear, but it can be recast in a non-linear form if
the constraints are taken into account. In this sense, the Hamiltonian
contains interactions.

\endchapter

\header{7. Discussion }

As stated in the introduction, our primary goal is to construct irreps
of $Vect(N)$ (with anomalies).  To achieve this goal, we start from a
set of modules which we understand well, (tensor and conformal
fields), and construct the corresponding Weyl algebras.  These are of
course huge modules, but we may hope that they contain irreducible
components which can be isolated by factoring out certain relations.
In this paper we have employed several methods of reduction, namely

\point Invariants. A single Fourier component of a molecule is set to
zero. This mechanism is responsible for quantization of the parameters
$\lambda$ and $w$.

\point Constraints. The fields are restricted to the solutions of
certain equations, involving all Fourier components.

\point Vacuum. A Fock module is much smaller than its parent Weyl
algebra, because all strings containing oscillators of negative degree
are eliminated. Na\"\i vely, a vacuum leads to infinities, which is
taken care of by normal ordering and imposing constraints.

\point Central charges. Any operator which commutes with all of
$Vect(N)$, e.g.  the fermionic number operator, has a single value in
an irrep. Hence any module decomposes into submodules of fixed value.

\point Gauge orbits. If some field admits a nontrivial action of a
gauge algebra, the $Vect(N)$ module decomposes into an infinite direct
sum of isomorphic copies.  Each copy can be considered as a
representation on a gauge orbit. Equivalently, we can enlarge the
algebra to $Vect(N) \IX Map(N, \oj)$, which is indeed done in
{\it rational} conformal field theory.  Representations on gauge orbits are
also closely related to the theory of Hamiltonian systems with
first-class constraints.\refto{21}

We believe that this programme to a large extent has been carried out
in this paper, at least conceptually, but many problems of technical
nature remain. The major difficulty lies perhaps in the non-linearity
of the constraints, which prevents us from writing down the general
solution as a sum of eigenmodes.  Another problem is to decide which
fields to start from. Mathematically, the ideal choice would be the
minimal set of fields from which all irreps can be built, but it is
not obvious what this is. It would also be desirable to have a more
concrete description of a single nontrivial irrep.  Finally, we should
of course check whether the modules are unitary, but there seems to be
little point in doing so before we know if they are irreducible.

Although the problem in this paper is purely mathematical, we expect
that it has consequences for physics. Our original reason for studying
$Vect(N)$ was to extend the classification of critical exponents to
$N$ dimensions. This goal now seems closer, because exceptionally
small modules can only be constructed for a discrete set of the
parameters, as was explained in Sec. 4.

However, the most striking result is the appearence of algebraic
structures similar to quantum gravity. If we by the word ``quantum''
mean that there is a lowest-weight state and by the word ``gravity''
mean that a conformal analog of the Einstein equation is satisfied,
then the modules in the previous section qualify as ``quantum
gravity''. Moreover, they are presumably consistent in the sense that
there are only finite anomalies. Of course, this is not what is
usually meant by quantum gravity. On the other hand, conformal
equations could not have been written down before conformal fields
were discovered, so this option has never been tested before. In any
event, we know of no other route to quantized gravity which is not
manifestly inconsistent, with the possible exception of string theory.

We can evidently build some kind of classical gravity theory using
conformal fields, as we indicated in Sec. 6.  It is not clear to us if
such a theory will differ in any significant way from the standard
Einstein theory for large time scales, compared to the Planck time.
It should in this connection be noted that $N$-dimensional conformal
fields in many respects are similar to $(N+1)$-dimensional tensor
fields, because vectors have an equal number of components.

Finally, we want to emphasize that all physical objects, including
those involved in quantum gravity, must transform consistently under
arbitrary coordinate transformations and hence they must be $Vect(N)$
representations. We find it plausible that Nature would use
irreducible representations as her fundamental building blocks.

\endchapter

\header{ Appendix}

The proof that (6.1) satisfies $Vect(N)$ is given here for convenience.
$$ \eqalign{
[L^\mu(m), &L^\nu(n)]
= \bigl[ \emx \bigl(\d^\mu + (m^A + k^A) T^\mu_A
+ c m^\mu m^A T^B_A x_B \bigr), \cr
&\qquad \enx \bigl(\d^\nu + (n^C + k^C) T^\nu_C
+ c n^\nu n^C T^D_C x_D \bigr) \bigr] \cr
&= \emnx \Big( n^\mu(\d^\nu + (n^C + k^C) T^\nu_C
+ c n^\nu n^C T^D_C x_D)
- n^\mu k^C T^\nu_C \cr
& \qquad + c n^\nu( - n^\mu k^C T^D_C x_D + n^C T^\mu_C)
+ (n^\mu + k^\mu) (m^A + k^A) T^\nu_A  \cr
&\qquad + c n^\nu (m^A + k^A) (n^\mu T^D_A x_D - n^C T^\mu_C x_A) \cr
& \qquad+ c^2 m^\mu n^\nu m^A n^B x_B T^D_A x_D \Big) - \mxn \cr
&= n^\mu \emnx \Big(\d^\nu + (m^C + n^C + k^C) T^\nu_C
+ c n^\nu (m^C + n^C) T^D_C x_D \Big) - \mxn \cr
&= n^\mu L^\nu(m+n) - \mxn. }
	\eqno(A.1) $$
We used that $m^A x_A = n^A x_A = k^\mu = 0$ and that
$$ n^\mu n^\nu f(m+n) - \mxn =  n^\mu (m^\nu + n^\nu) f(m+n) - \mxn
	\eqno(A.2) $$
for any expression that depends on the sum $m+n$ only.

\endchapter
\newpage

\noindent\header{References}

\noindent 1.
A. A. Belavin, A. M. Polykov and A. B. Zamolodchikov,
Nucl. Phys. B 241 (1984) 333

\noindent 2.
C. Itzykson, H. Saleur and J.-B. Zuber,
Conformal invariance and applications to statistical mechanics,
World Scientific, Singapore (1988)

\noindent 3.
D. J. Amit, Field Theory, the Renormalization Group, and Critical
Phenomena, McGraw-Hill (1978)

\noindent 4.
T. Eguchi, P. B. Gilkey and A. J. Hansson, Phys. Rep. 66 (1980) 213

\noindent 5.
C. Nash and S. Sen, Topology and geometry for physicists,
 Academic Press, London (1983)

\noindent 6.
E. Ramos and R. E. Shrock, Int. J. Mod. Phys. A 4 (1989) 4295

\noindent 7.
E. Ramos, C. H. Sah and R. E. Shrock, J. Math. Phys. 31 (1989) 1805

\noindent 8.
F. Figueirido and E. Ramos, Int. J. Mod. Phys. A 5 (1991) 771

\noindent 9.
T. A. Larsson, Phys. Lett. B 231 (1989) 94

\noindent 10.
 T. A. Larsson, J. Phys. A 25 (1992) 1177

\noindent 11.
T. A. Larsson, to appear in Int. J. Mod. Phys. A (1992)
(hep-th/9207029)

\noindent 12.
T. A. Larsson, submitted (1992)
(hep-th/9207030)

\noindent 13.
B. L. Feigin and D. B. Fuks, Funct. Anal. and Appl. 16 (1982) 144

\noindent 14.
E. Date, M. Jimbo, M. Kashiwara and T. Miwa, Proc. of RIMS symp.,
ed. M. Jimbo and T. Miwa, World Scientific, Singapore (1983) 39

\noindent 15.
A. Tsuchiya and Y. Kanie, Publ. RIMS, Kyoto Univ. 22 (1986), 259

\noindent 16.
P. Goddard and D. Olive, Int. J. Mod. Phys. A 1 (1986) 303

\noindent 17.
J. Mickelsson, Current algebras and groups, Plenum Press,
New York (1989)

\noindent 18.
J. Mickelsson and S. Rajeev, Commun. Math. Phys. 116 (1988) 365

\noindent 19.
J. Mickelsson, preprint (1992)

\noindent 20.
A. Pressley and G Segal, Loop groups, (Oxford: Claredon Press 1986)

\noindent 21.
P. A. M. Dirac, Lecture on Quantum Mechanics, Yeshiva University
 Academic Press, New York (1967)

\vfill
\end